\documentclass[fleqn,usenatbib]{mnras}

\usepackage{newtxtext,newtxmath}

\usepackage[T1]{fontenc}

\DeclareRobustCommand{\VAN}[3]{#2}
\let\VANthebibliography\thebibliography
\def\thebibliography{\DeclareRobustCommand{\VAN}[3]{##3}\VANthebibliography}

\usepackage{graphicx}	
\usepackage{amsmath}	

\usepackage{gensymb}
\usepackage[dvipsnames]{xcolor}
\definecolor{changes}{RGB}{0 , 0, 0} 

\title[Effect of single fly-bys on close-in planets]{The effect of dynamical interactions in stellar birth environments on the orbits of young close-in planetary systems}

\author[Schoettler, C. and Owen, J. E.]{Christina Schoettler$^{1}$\thanks{E-mail: c.schoettler@imperial.ac.uk} and James E. Owen$^{1}$
\\
$^{1}$Astrophysics Group, Department of Physics, Imperial College London, Prince Consort Rd, London SW7 2AZ, UK}

\date{Accepted XXX. Received YYY; in original form ZZZ}

\pubyear{2024}

\begin{document}
\label{firstpage}
\pagerange{\pageref{firstpage}--\pageref{lastpage}}
\maketitle

\begin{abstract}
Stars do not form in isolation but together with other stars, and often in a clustered environment. Depending on the initial conditions in these environments, such as initial density and substructure, the distances of encounters between stars will differ. These encounters can also affect just-formed exoplanetary systems. Using N-body simulations, we show the effect of a single fly-by on a common type of exoplanetary system: close-in Super-Earths/sub-Neptunes with or without a distant Giant planet. Even a single encounter can significantly modify the architecture of these exoplanetary systems over their long lifetimes. We test fly-bys with different characteristics, such as distance and mass, and show how they perturb the inner planets long after the encounter, leading to collisions and mutual inclination excitation, which can significantly modify the observed architecture of these systems in transit. We find that our initially four-planet inner systems reduce to three or two inner planets depending on their initial separation and that the mutual inclinations of these remaining planets can be high enough to reduce the number of observable, transiting planets. In our 500 Myr simulations, we show that this reduction in the number of transiting planets due to stellar fly-bys can contribute to the observed excess of single-transit systems.

\end{abstract}

\begin{keywords}
methods: numerical -- planets and satellites: dynamical evolution and stability -- planet–star interactions -- planetary systems
\end{keywords}



\section{Introduction} \label{Intro}

The birth environment of planets is intrinsically linked with that of their host stars, as planets form from the protoplanetary discs around these young protostars \citep[e.g.][]{2015ApJ...809...93D,2020Natur.586..228S,2020ApJ...904L...6A,2020ApJ...891..166W,2020RSOS....701271P,2023ASPC..534..685P}. Many of the exoplanet host stars have likely formed with other stars in grouped or clustered environments. While these regions are still young, stars and any planets around these stars will be subject to a higher-density environment than what they will eventually experience after dispersal into the field \citep[e.g.][]{RN25,RN59}. As such, this birth environment has the potential to influence not only the stars and their protoplanetary discs but also any planets that have already formed around them. 

The effects of dynamical interactions caused by being in an evolving young star-forming region have been investigated for different types of single \citep[e.g.][]{1998ApJ...508L.171L,2009ApJ...697..458S,2012MNRAS.419.2448P,2022MNRAS.514..920D} and multi-planet architectures \citep[e.g.][]{2011MNRAS.411..859M,2013MNRAS.433..867H,2016ApJ...816...59S,2017MNRAS.470.4337C,2019MNRAS.489.4311C,2019A&A...621A.101B,2019A&A...624A.120V,2020MNRAS.497.1807S,2022MNRAS.512.2460S,2020ApJ...905..136W,2022MNRAS.509.5253W,2023MNRAS.520..637R} as well as how planets still within their protoplanetary discs are affected \citep[e.g.][]{2014A&A...564A..28P,2023MNRAS.tmp.2714F}. Fly-bys have even been thought to have had an influence on the shape of our own outer Solar system \citep[e.g.][]{2018ApJ...863...45P,2021A&A...651A..38P,2019MNRAS.489.2280F,2022MNRAS.515.5942B}. A lot of this work has focused on more distant, colder Giant planets, however, observations indicate that they are rarer \citep[$\sim$10--20 per cent, e.g.][]{2008PASP..120..531C,2021ApJS..255...14F, 2022ApJS..262....1R}. Close-in planets that can be observed via transits are more common \citep[$\sim$30--50 per cent, e.g.][]{2013PNAS..11019273P,2018AJ....156...24M,2018ApJ...860..101Z, 2022ApJS..262....1R}, which is why we focus on the influence of an encounter on these smaller planets here.

Our knowledge about planet systems around other stars has dramatically evolved over the past two decades, and the most productive planet-finding mission to date was Kepler, which launched in 2009 with its primary mission lasting $\sim$4 years. It used the transit method to discover thousands of planets, both in single-transit and multi-transit systems \citep[e.g.][]{2010arXiv1001.2010W,2014ApJ...790..146F}. A majority of these discovered planets are smaller (1-4 R$_{\oplus}$), relatively low-mass planets (< 20 M$_{\oplus}$) in close-in orbits (period < 100 days). 

For a multi-planet system to be discovered by the transit method requires its planets to be orbitally aligned with respect to the observing direction, necessitating low mutual inclinations. However, even in a fully coplanar configuration, not all close-in planets are guaranteed to transit their host star together \citep[e.g.][]{2010arXiv1006.3727R}. The Kepler mission discovered about three times as many single-transiting systems compared to the number of systems with more than one transiting planet. \citet{2011ApJS..197....8L} showed that this difference in the numbers of transiting planets towards more single-transit systems was likely not the result of a single underlying planet population of low mutual inclination systems ($\sim$1\degree--2\degree) with the same initial number of planets, as this would result in too few single-transit systems. Shortly thereafter, \citet[][]{2012ApJ...758...39J} referred to this observational excess as a dichotomy between single-transit and multi-transit systems and it has since been known as the ``Kepler Dichotomy''. 

\citet{2019MNRAS.483.4479Z} investigated the Kepler detection pipeline of multiples to determine if the dichotomy was actually a real feature instead of a detection bias. They found that incompleteness caused the dichotomy to probably be exaggerated but failed to account for it fully. Regardless of this aspect, the dichotomy is now generally assumed to be due to differences in the inclination of planets in the same system. However, it is unclear if this is due to the existence of a population of planets with higher mutual inclinations in addition to a lower-inclination population or just a spread of inclinations with a high inclination tail from a single population \citep[e.g.][]{2018AJ....156...24M, 2019MNRAS.490.4575H,2020AJ....160..276H,2022arXiv220310076W}. In particular, \citet{2019MNRAS.490.4575H} showed that the dichotomy was likely not due to a population of single-planet systems. \citet{2021AJ....162..166M} showed that mutual inclinations are distributed continuously and suggested that this result pointed to the oblateness of the host star or late assembly of the planets as the cause for the dichotomy. 

While the inclination-population model explains why we might observe the dichotomy, it does not provide an underlying cause for the different inclinations. The explanations brought forward for the origin of these differences can be split into two different avenues. The first one comprises theories based on the idea that differences in/during planet formation lead to the observation of a dichotomy. 

\citet{2012ApJ...758...39J} suggested that giant planet-planet scattering or migration during the formation process prevented the subsequent formation of smaller, close-in planets, leaving highly inclined large planets behind. They noted that single-transiting systems did not seem to have a preferred planet size, whereas multi-transiting systems appeared to preferentially host smaller planets. Alternatively, \citet{2016ApJ...832...34M} proposed that differences in the protoplanetary disc surface density and mass could result in differences in, e.g. multiplicity and mutual inclination. Another explanation is centred around a misalignment of the spin-orbit of the star with its planets as a consequence of excitation during the disc phase. This then caused subsequent instabilities in the planetary orbits and higher mutual inclination differences \citep[e.g.][]{2016ApJ...830....5S}. Simulations by \citet{2017MNRAS.470.1750I,2021A&A...650A.152I} suggested that instabilities in the resonant chains created during the formation of the planets were another possible cause. 

\citet{2022ApJ...937...53Z} found that the dichotomy could be caused not by higher mutual inclinations of close-in planets but by fewer planets within the Kepler detection window. This was thought to be due to differences in migration and clustering of planets trapped outside the transit-observable inner au-region. In addition to this, the mass of the discs was found to also play a role in how many of these planet clusters would form.

The second explanation route for the dichotomy is based on the idea that inclination differences are caused by the long-term evolution of the planet systems in question. This evolution can either be investigated in isolation or with other stars impacting the planetary systems. There appears to be a correlation between the occurrence of inner Super-Earth planets and a more distant outer giant \citep[e.g.][]{2016ApJ...821...89B,2019AJ....157...52B,2018AJ....156...92Z,2021A&A...656A..71S}. While the distant giant planet would not be detected in transit surveys, it could nevertheless influence the dynamical evolution of the inner planet system. This is the idea behind several studies investigating the dynamical evolution of a planet system with an outer giant. Perturbations of the inner planets' orbits can be caused by the presence of an outer giant planet (system), e.g. when it is on an inclined/misaligned orbit or \textcolor{changes}{ becomes dynamically unstable \citep[e.g.][]{2017AJ....153...42L,2017MNRAS.467.1531H,2017MNRAS.469..171R,2017MNRAS.468.3000M,2018MNRAS.478..197P,2019MNRAS.482.4146D,2020MNRAS.498.5166P}}. 


Interactions with other stars in the birth environment were suggested as a cause of disruption in a planet system, potentially explaining the dichotomy. \citet{2018MNRAS.474.5114C} concluded from their simulations of the evolution of planet systems in a young star cluster that single-transit systems could originate from denser regions within the birth environment. \citet{2020MNRAS.496.1149L} investigated the effect of a fly-by on different star-planet systems, each with two planets in different configurations. While these authors did not use their simulations to investigate the dichotomy directly, they found that fly-bys of their close-in SE systems around an M-dwarf could not disrupt them sufficiently to depart from their initial coplanar set-up. This result could be in part due to their relatively short integration time of only 100 Myr.

\citet{2022MNRAS.509.1010R} also investigated if stellar fly-bys could cause excitation of initially coplanar planetary systems to explain the dichotomy. They found that the presence of one or more exterior larger companion planets decided the amount of disturbance in the inner planet system after a fly-by. Their idea was centred around the giant planets being disturbed by the fly-by, which then caused the inner planet system to destabilise. However, they only evolved their planetary systems for a short period of time (<1 Myr) after the fly-by occurred, not considering that the perturbations introduced by the fly-by could take several Myr to hundreds of Myr to have any measurable disturbing effect \citep[see e.g.][]{2011MNRAS.411..859M}.

In this paper, we take a similar approach of simulating the effect of a single stellar fly-by on a planetary system that could be observed using the transit method, but we follow the evolution of the planet systems for 500 Myr after the encounter. The fly-by characteristics are based on typical stellar encounters within a dynamically evolving young star-forming region that we extracted from $N$-body simulations of a typical region (based on the Orion Nebula Cluster - ONC). Our planetary systems either feature a distant giant planet around a close-in system of smaller planets or only of the close-in system. In section \ref{method}, we describe the close encounters that can occur in a typical young star-forming region during its early dynamical evolution and the type of fly-bys we investigate in our simulations. We then provide details about the planetary systems that are subjected to a fly-by and how we determine the number of transiting planets. In section \ref{results}, we provide the results of our planet system simulations and discuss them in section \ref{discussion}. Finally, section \ref{conclusion} provides concluding remarks.

\section{Method}\label{method}

\subsection{Determining typical interactions between stars}

We simulate the effect of typical interactions of young star-planet systems with other stars in young star-forming regions. The input for these interactions (i.e. perturber mass, fly-by distance, and perturber velocity) are based on close encounter information from 20 simulations of the dynamical evolution of a typical (Orion Nebula Cluster - like) star-forming region used in \citet{2020MNRAS.495.3104S}. These $N$-body simulations simulated the early dynamical evolution of stars (without planets) in young star-forming regions. They are set up using the box fractal method \citep{RN14}, which can create different levels of initial kinematic and spatial substructure to mimic what has been found in observations. A detailed overview of how fractals can be used to create substructure in simulations and how to construct them can be found in \citet{RN14} and also \citet{RN5, RN1}. For the stellar interactions, we choose simulations that start with a fractal dimension of $D$ = 2.0, which represents a high amount of initial spatial substructure and an initially subvirial ratio ($\alpha_{\text{vir}}$ = 0.3), which sets up the initial kinematic substructure. The virial ratio is the ratio of the total kinetic energy to the total potential energy of all stars in the simulation.  These substructures and subvirial initial conditions are representative of those that have been suggested for typical star-forming regions, like the Orion Nebula Cluster \citep[e.g.][]{RN4, RN38,2020MNRAS.495.3104S}.

\begin{figure}
        \centering
        \vspace{0pt}
    	\includegraphics[width=1.0\columnwidth]{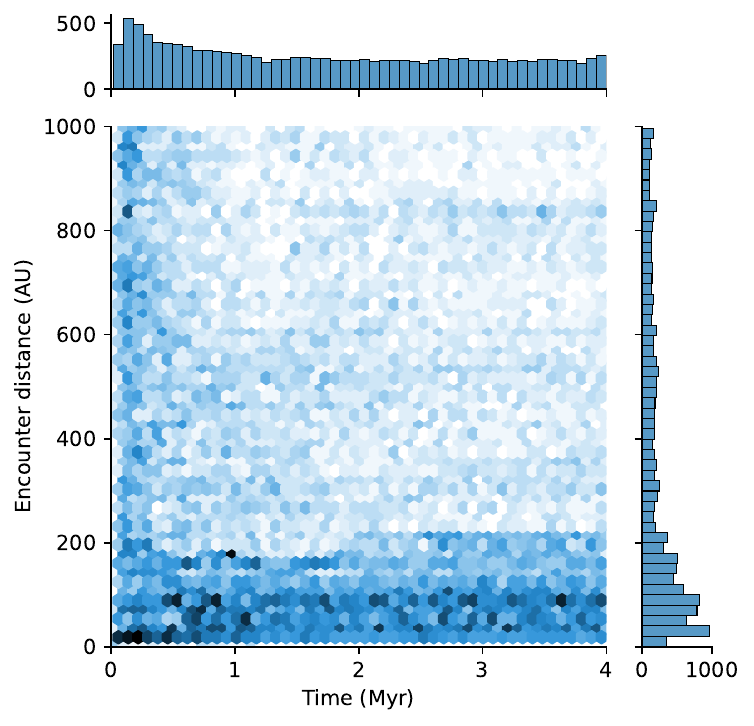}
    \caption{Example showing the close encounter distances of all non-primordial binary interactions from a single simulation with ONC-like initial conditions against the 4 Myr simulation time \citep[simulations from][]{2020MNRAS.495.3104S}. The interactions are more concentrated during the early evolution of the region (< 0.5 Myr) during which these regions undergo cool collapse. During the later part of the simulation, the number of interactions per timestep levels out and continues at around the same occurrence rate. The number of bins on both axes is 50 resulting in bins with a size of 80 kyr on the x-axis and 20 au on the y-axis}
    \label{fig:examples}
\end{figure}

These simulations include primordial binaries as well as stellar evolution. The number of systems is 2000, with the primary masses sampled randomly from a \citet{RN203} IMF. This IMF is a combination of a Chabrier (\citeyear{RN200}) lognormal IMF approximation for lower-mass stars with the power-law slope of Salpeter (\citeyear{RN204}) for stars more massive than 1 M$_{\sun}$. The presence of primordial binaries increases the number of stars to $\sim$2800 per simulation. The primaries and single stars have masses within a range from 0.1 M$_{\sun}$ and 50 M$_{\sun}$, whereas the secondary binary components can have masses down to 0.01 M$_{\sun}$ (i.e. brown dwarfs). Further details, e.g. on setting up the stellar and binary populations in the simulations, can be found in \citet{2020MNRAS.495.3104S}.

Fig.~\ref{fig:examples} shows the distribution of the (non-primordial binary) close interaction distances plotted against simulation time (in Myr) from one of these 20 previously described simulations. Star-forming regions, such as the ONC-like ones simulated in \citet{2020MNRAS.495.3104S} and used here, undergo ``cool'' collapse due to their initially substructured and subvirial nature \citep[e.g.][]{RN4,RN1}. This evolution can occur on very short timescales, which is evident in numerous close interactions within the first 1 Myr. 

\begin{figure}
        \centering
        \vspace{0pt}
        \includegraphics[width=0.9\columnwidth]{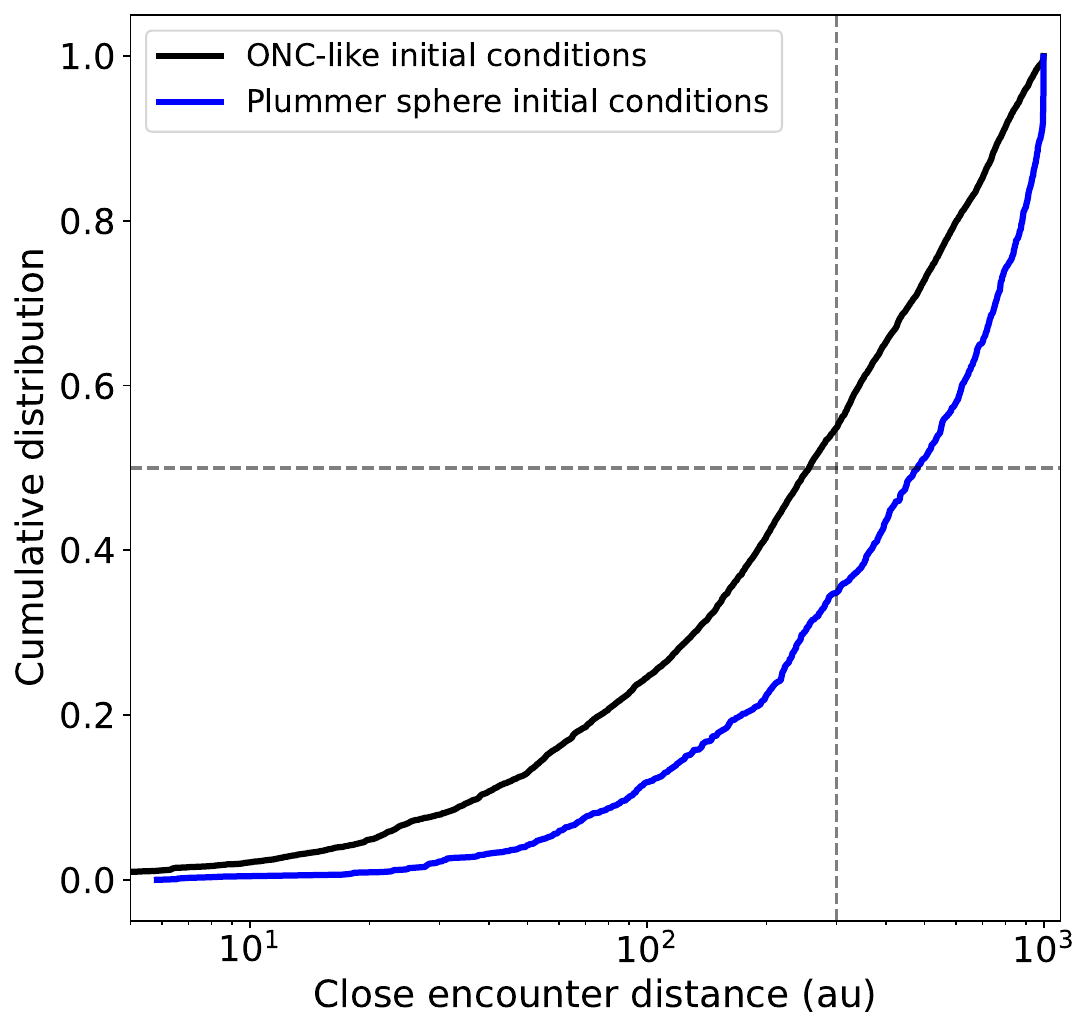}
    \caption{Comparison of cumulative distributions of close interactions \textcolor{changes}{where the encountering stars are not part of the same primordial binary, i.e. excluding binary encounters. but not encounters of binary component stars after they become unbound from their primordial binary. This data is taken as an average from 20 simulations each set up with two different initial conditions \citep[ONC-like simulations ('black') and a Plummer sphere ('blue'),][]{2020MNRAS.495.3104S} during 1--4 Myr simulation time, which is after the simulated star-forming regions have undergone their densest core-collapse phase. The Plummer sphere \citep{RN198} simulations shown here were set up initially virialised without primordial binaries or stellar evolution and a half-mass radius of 0.2 pc and do not undergo core-collapse. For the ONC-like initial conditions, more than half of these interactions occur below a distance of 300 au (dashed 'black' lines).}}
    \label{fig:percentage}
\end{figure}

\textcolor{changes}{The encounters considered here do not include those that occur between two stars that are in a primordial binary but do include those between the binary stars and other stars. More than half of the encounters occurring after 1 Myr (after core-collapse) have a close encounter distance <300 au. This encounter distribution is evident in Fig.~\ref{fig:percentage}, where we show a cumulative distribution of encounter distances between 1-4 Myr of simulation time from 20 simulations combined for two different initial condition sets. Differences between the physically more realistic ONC-like and the often used Plummer sphere \citep{RN198} initial conditions are in the number of stars: $\sim$2800 stars including primordial binaries versus 1000 stars without primordial binaries. The Plummer sphere simulations were set up without stellar evolution, however, this is expected to have little effect on the encounter profiles over the 4 Myr evolution time considered here and they are initially virialised. The ONC-like simulations (initially subvirial and spatially substructured) have similar encounter profiles compared to those fitting other star-forming regions \citep[e.g. NGC 2264, simulations from][]{2022MNRAS.510.3178S}. The Plummer sphere distribution is distinctly different, with encounters typically occurring at larger separations i.e. 50 per cent are larger than $\sim$500 au.}
For our single fly-by analysis, we select three different distances (50 au, 150 au and 250 au) \textcolor{changes}{covering the distance below which half the encounters occur (300 au) }in the star-forming region simulations. In our fly-by simulations, these close encounter distances are set in relation to the centre of mass of the star-planet system. 

\textcolor{changes}{In the 20 simulations (ONC-like initial conditions) after core-collapse during 1-4 Myr of the simulation time,  up to 23.4 per cent of the initially single, approximately solar-mass (0.9-1.1 M$_{\sun}$) stars have at least one encounter with another star/system at a distance <1000 au and up to 12.5 per cent have at least one encounter with another star at <300 au. In contrast, up to 2.2 per cent of single stars with masses less than 0.9 M$_{\sun}$ and up to 19.8 per cent of those with masses above 1.1 M$_{\sun}$ have at least one encounter below <300 au}. For the \textcolor{changes}{following} star-planet simulations, we chose three different perturber masses (0.1, 0.5 and 1.0 M$_{\sun}$), whereas the star hosting the planet systems is always a solar-mass star (1.0 M$_{\sun}$).

\begin{table*}
	\centering
	\caption{Input parameters for the fly-by simulations. Column 1: Planetary systems only containing 4 Sub-Neptunes/Super-Earths. Column 2: Planetary systems containing 4 inner planets (SNs/SEs) and an outer GP.}
	\label{tab:Rebound_sim_para}
	\begin{tabular}{lcc}
		\hline
		 & 4 inner planets only &  4 inner planets + 1 giant planet \\
		\hline	
		Inclination incl$_{\text{pl}}$ & random  & random  \\ 
		Inner planet separation ($R_{\rm{HM}}$)  & 10, 14 &  10, 14   \\ 
		Giant Planet semi-major axis (au) & - &  5, 10, 20  \\
            Fly-by distances (au) & 50, 150, 250 & 50, 150, 250 \\
            Fly-by velocity (km\,s$^{-1}$) & 2, 4 & 2, 4\\
            Fly-by stellar mass (M$_{\sun}$) & 0.1 , 0.5, 1.0 & 0.1 , 0.5, 1.0 \\
        \hline
	\end{tabular}
\end{table*}

Finally, we select two different perturber velocities based on the median velocities of the non-binary stars in the stellar simulations. The median stellar velocities across the 20 simulations are very similar at $\sim$2 km\,s$^{-1}$. We assume two different fly-by cases, first the extreme case of two stars passing each other in exactly opposite directions, each with this median velocity. In a reference frame centred on the host star, the velocity of the perturber would be double, i.e. $\sim$4 km\,s$^{-1}$. The second perturber velocity we test in the fly-by simulation uses half of this velocity, simulating a slower fly-by ($\sim$2 km\,s$^{-1}$). The perturbers start their journey at distances between $\sim$8000--9000 au at different positions in relation to the simulated planetary systems. The perturbers take around $\sim$9,000--10,000 and $\sim$18,000--20,000 yr (depending on their velocity and close encounter distance) to reach the host star and interact with the planetary system located there.

\subsection{Setting-up and evolving the planetary systems}\label{Method_planets}

As we are interested in investigating whether close encounters in young star-forming regions can play a part in explaining the Kepler dichotomy, we set up two different exoplanet architectures that are commonly associated with those found via the transit method. These architectures all feature four close-in planets, each with a mass of 5 M$_{\oplus}$ around the 1.0 M$_{\sun}$ host star. \textcolor{changes}{This choice of using close-in planets with initially the same mass is motivated by the results in \citet{2017ApJ...849L..33M}, who showed intra-system uniformity in planet masses in multi-planet systems (similar to the intra-system uniformity seen in exoplanet radii \citealt{2018AJ....155...48W}).}

Planets with these masses are classified as Super-Earths (SE) or Sub-Neptunes (SN), depending on their composition. As we use fly-by information from the early dynamical evolution of a star-forming region, we assume that these planets are still SNs with a gaseous atmosphere, resulting in higher radii than if they were SEs. We chose a radius of three R$_{\oplus}$ as planets at this young age are more likely to have a large H/He atmosphere/envelope \citep[e.g.][]{2013ApJ...776....2L,2014PNAS..11112655M,2021MNRAS.503.1526R}. 

The radii of the stars (host and perturber) are chosen based on their mass using a main-sequence mass-radius relationship. We disregard that at this early stage of the dynamical evolution of a star-forming region, they might still be pre-main sequence stars with radii larger than what they would have once they reached the main sequence \citep[e.g.][]{2003RPPh...66.1651L}. The only effect that a larger solar radius would have in our simulations is to cause collisions of planets with the host star (which do not occur in our simulations). All solar-mass stars in our simulation have a radius of 1 R$_{\sun}$ and the two sub-solar mass perturbers (0.5 M$_{\sun}$ and 0.1 M$_{\sun}$) have radii following the mass-radius-relationship R $\propto$ M$^{0.8}$. 


We do not evolve the planets' radii unless they are involved in a collision. If two SNs collide with one another, the gaseous envelope is likely removed, leaving a rocky SE behind, as suggested by the result of giant impacts in \citep[e.g.][]{2015ApJ...812..164L,2019MNRAS.485.4454B,2023ApJ...954..196K}. To calculate the new radius, we use the mass-radius relationship for rock/iron planets based on Eq. 8 in \citet{2007ApJ...659.1661F,2007ApJ...668.1267F} with a rock to iron fraction rmf = 2/3, i.e. Earth-like composition. Our collisions result in a perfect merging of the two involved planets, with their masses combining to form a larger planet. 


The inner-most planet is always placed at 0.1 au from the host star, which is a location commonly associated with Kepler close-in planets \citep[e.g.][]{Petigura2022}. The other three planets are placed at a separation based on their mutual Hill radius $R_{\rm{MH}}$. The mutual Hill sphere can be calculated using
\begin{equation}\label{Hill_sphere}
R_{\rm{MH}} = \cfrac{a_1 + a_2}{2} \left( \cfrac{m_1 + m_2}{3 M} \right)^{1/3},
\end{equation}
where m$_i$ and a$_i$ are the masses and semi-major axes of two adjacent planets and the central host stellar mass is $M$. The above calculates a radius, which can be put in relation to the separation between these two planets to calculate a dimensionless number K:
\begin{equation}\label{Hill_K}
K = \cfrac{a_2 - a_1}{R_{\rm{MH}}} 
\end{equation}
\citet{2015ApJ...807...44P} suggested that planetary systems with similar masses set up in circular and coplanar orbits and a minimum of 10 $R_{\rm{MH}}$ are stable for at least 10$^9$ yr. \citet{2018AJ....155...48W} also found that 93 per cent of their observed planet sample had separations with at least this K-value. 
We set up our initially circular and coplanar planetary systems with separations of the inner planets of 10 and 14 $R_{\rm{MH}}$. The former is just at the edge of stability, whereas the latter is suggested to be stable even if eccentricities are non-zero \citep[see][]{2015ApJ...807...44P}. We run simulations of these systems in isolation (without perturber), and all of them remain virtually coplanar (mutual inclination < 10$^{-7}\degree$) and circular (eccentricity < 10$^{-3}$) up to at least 500 Myr, which is the age which we run all of our fly-by simulations to in this analysis. This simulation time is much shorter than the average ages of many known Kepler planet systems, but it is not feasible to run our systems for this long, e.g. 5 Gyr or more.

As mentioned in section~\ref{Intro}, there is an apparent connection between the occurrence of close-in planets with a distant Giant \citep[e.g.][]{2016ApJ...821...89B,2019AJ....157...52B,2018AJ....156...92Z,2021A&A...656A..71S}. In our second set-up, we add simulations where the already described inner planetary systems feature a more distant outer Giant with a mass of 5 M$_{\rm{J}}$ at three different distances to the host star (5, 10 and 20 au). We set the radius of this Giant planet to 1 R$_{\rm{J}}$ \citep[e.g.][]{2011ApJS..197...12D,2019RNAAS...3..128T}.

For all of the planetary architecture combinations, we simulate two different orbital starting positions for the inner planets. They are either lined up at the start or separated by a quarter of their orbital period. The inclinations of the planetary systems as a whole are chosen at random; mutual inclinations are zero. For each of our 18 perturbers (differing in mass [3 options], velocity [2 options] and close encounter distance [3 options]), we use two different starting positions, resulting in a total of 36 different fly-bys. In addition to the random choice of planet system inclinations, this last choice results in different interaction angles between the perturbers and planets for otherwise identical simulations.

We use the REBOUND N-body code \citep{rebound} to integrate the systems. We run a total of 576 fly-by simulations using different planetary orbit and fly-by configurations. The input parameters for these simulations are summarised in Table\,\ref{tab:Rebound_sim_para}.


\begin{figure*}
 \centering
         \centering
        \vspace{0pt}
    	\includegraphics[width=0.9\textwidth]{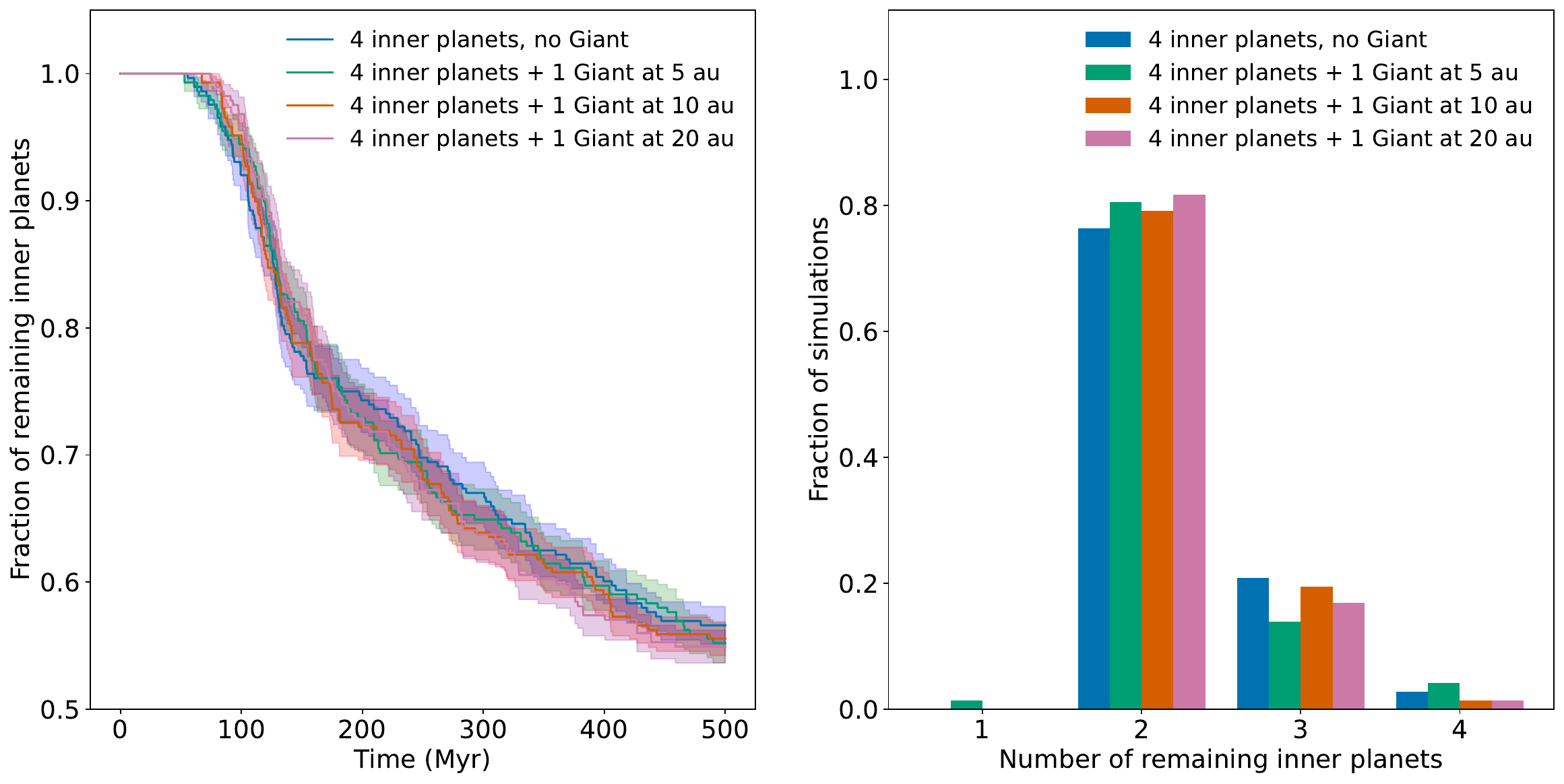}
     \caption{\textbf{Left:} Evolution of the average number of inner planets over the 500 Myr simulation time for the 10 R $R_{\rm{MH}}$ simulations shown as a fraction of the number of original inner planets. The first collisions between the inner planets start at $\sim$53 Myr in the simulations with a Giant at 5 au. The average numbers for all four cases then drop off and reach an average value of $\sim$2.2--2.3 planets per system at 500 Myr (approx. half of the original number). The drop and value of the average number of planets in all four architecture set-ups are very similar over time and also start before 100 Myr simulation time for all of them. The coloured shaded regions show the standard error of the mean and indicate the spread of the values of the number of planets in different simulations. \textbf{Right:} Histogram of the number of remaining inner planets at 500 Myr for the four set-ups shown as a fraction of simulations. Starting with four inner planets in all simulations, most systems ($\sim$80 per cent) lose at least two of them due to collisions. In one simulation, all inner planets collide with each other, leaving only one inner planet; none lose all their inner planets in collisions (there are no ejections of inner planets in any of our simulations).}
     \label{fig:Hist_planet_10}
 \centering
        \centering
        \vspace{0pt}
    	\includegraphics[width=0.9\textwidth]{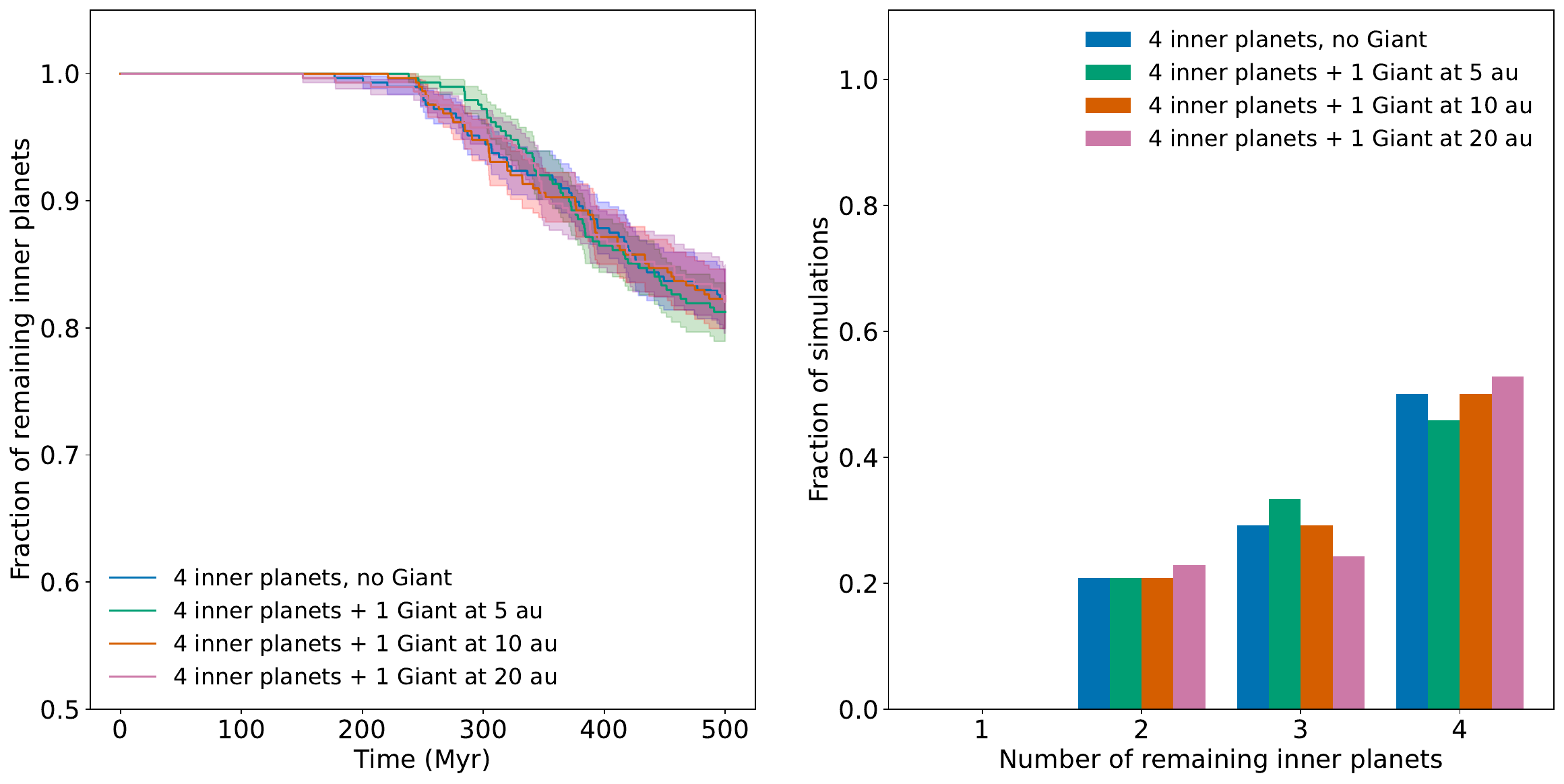}
     \caption{\textbf{Left:} Evolution of the average number of inner planets over the 500 Myr simulation time for the 14 R $R_{\rm{MH}}$ simulations shown as a fraction of the number of original inner planets. The first collisions between the inner planets start much later than in 10 R $R_{\rm{MH}}$ simulations at $\sim$160 Myr. The average numbers for all four cases then drop less rapidly and reach an average value of $\sim$3.3 planets per system at 500 Myr ($\sim$0.7 times the original number). The impact of the position of the Giant planet in relation to the start of collisions seems to favour an earlier disruption of the inner planet systems either without a Giant or one at 20 au. However, the coloured shaded regions show the standard error of the mean and their overlap indicates no statistical significance in the differences in the mean between architectures. \textbf{Right:} Histogram of the number of remaining inner planets at 500 Myr for the four set-ups shown as a fraction of the simulations. Starting with four inner planets in all simulations, half of the systems keep all of their planets during the simulation time, and the remainder of the simulations lose either one or two planets to collisions. We have no simulations where all inner planets collide with each other leaving only one inner planet.}
     \label{fig:Hist_planet_14}
\end{figure*}

We run these simulations up to a simulation time of 500 Myr. The fly-by occurs early (first 20 kyr) in the simulation, but we continue the integration with the perturber up to a simulation time of 1 Myr in the first instance. The simulation snapshots are stored in the REBOUND SimulationArchive with timesteps of 10$^3$ yr during this period using the IAS15 15th order Gauss-Radau integrator \citep{reboundias15,reboundsa}. To speed up the simulation, the perturber is removed at this time (1 Myr). We also change the integrator to the hybrid MERCURIUS integrator \citep{reboundmercurius} and increase the time between stored snapshots to 10$^4$ yr. 

The MERCURIUS integrator is a combination of two different integrators. When the particles (e.g. planets) are far apart from each other, it uses the symplectic Wisdom-Holman integrator WHFast with a fixed integration time step calculated from the period of our inner-most planet multiplied by 10$^{-2}$ \citep{2015MNRAS.452..376R}. For most of our simulations, this results in an internal fixed integration timestep dt$\approx$0.12 days based on the period of $\sim$11.6 days of the inner planet located at 0.1 au. 

The hybrid MERCURIUS integrator automatically switches over to the high-order IAS15 integrator during close encounters. We set a generous switch-over distance at a value of 5 $R_{\rm{MH}}$, which is half the smaller initial separation distance of the IP systems. In this way, IAS15 is not triggered while the planets remain at their original separations. To prevent IAS15 from stalling during very close encounters, we set a minimum timestep of 10$^{-4}$\,dt \citep{reboundmercurius}.

\begin{figure*}
 \centering
         \centering
        \vspace{0pt}
    	\includegraphics[width=0.9\textwidth]{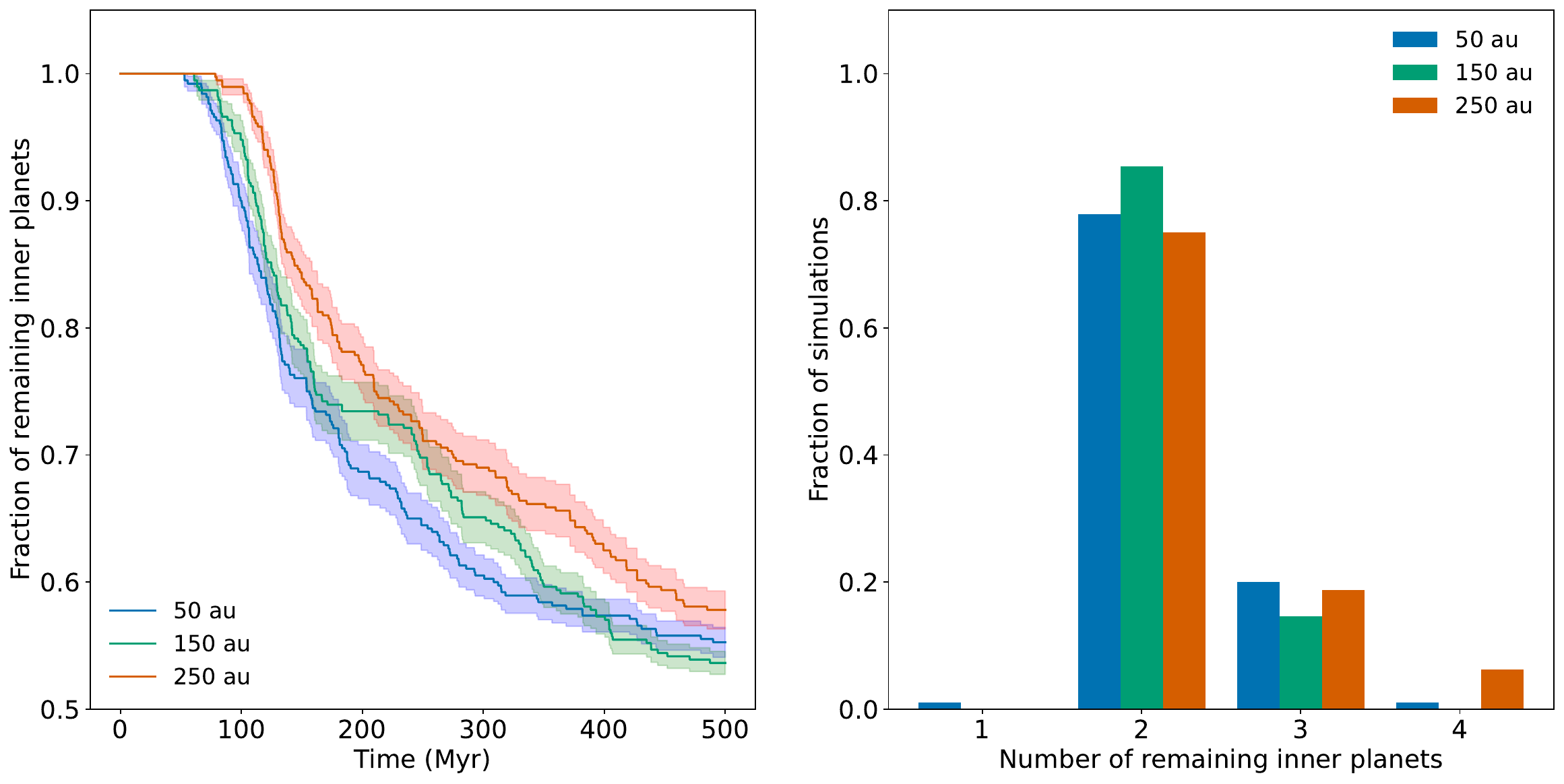}
     \caption{\textbf{Left:} Evolution of the average number of inner planets over the 500 Myr simulation time for the 10 $R_{\rm{MH}}$ simulations. Collisions between the inner planets start for the planet systems with a 50 au fly-by first ('blue') followed by the 150 au ('green') and 250 au ('red') fly-bys. The average numbers for all three cases drop off rapidly after the collisions commence and reach an average value of $\sim$2.2--2.3 planets per system at 500 Myr (approx. half of the original number). This drop in the average number of planets for all three fly-by distances is very similar over time. However, the coloured shaded region showing the standard error of the mean shows that there is a difference in the average values for most of the simulation times with different fly-by differences. \textbf{Right:} Histogram of the number of remaining inner planets at 500 Myr for the three distances as a fraction of simulations. Starting with four inner planets in all simulations, most systems lose at least two of them due to collisions, with the 150 au simulation showing the largest number of simulations with only two remaining inner planets and none of the systems subject to this fly-by distance retaining their original four inner planets. However, only one of our simulations finishes with one remaining inner planet (50 au fly-by) and none lose all these planets to collisions.}
     \label{fig:Hist_planet_10_2}
 \centering
         \centering
        \vspace{0pt}
    	\includegraphics[width=0.9\textwidth]{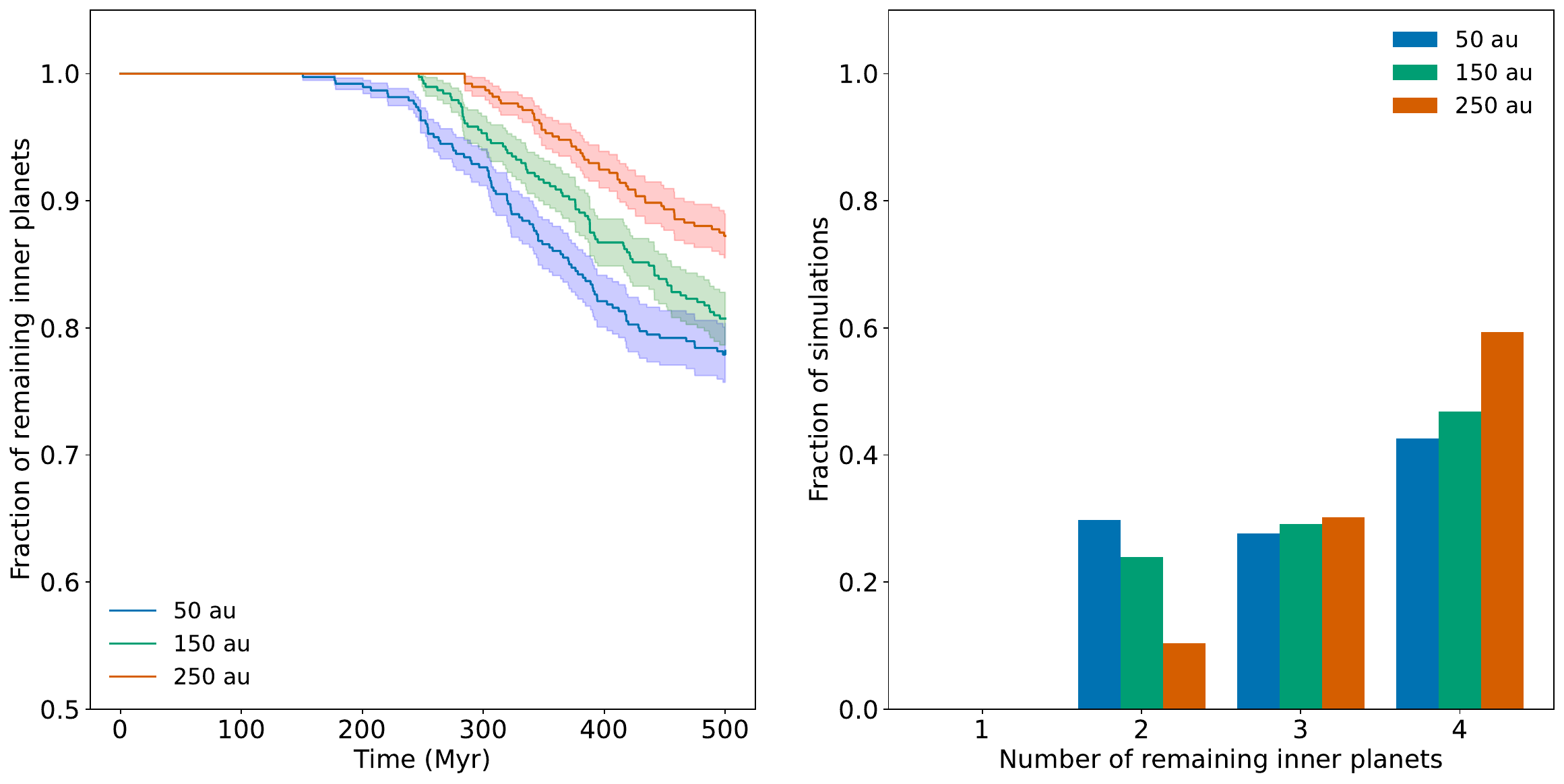}
     \caption{\textbf{Left:} Evolution of the average number of inner planets over the 500 Myr simulation time for the 14 $R_{\rm{MH}}$ simulations. The first collisions between the inner planets start much later than in 10 $R_{\rm{MH}}$ simulations at $\sim$160 Myr and then are staggered by increasing fly-by distance. The average numbers for all three cases drop less rapidly and reach an average value of $\sim$3.3--3.4 planets per system at 500 Myr ($\sim$0.8--0.9 times the original number). The drop in the average number of planets for all three fly-by distances is more pronounced for the systems subjected to closer fly-bys. The shaded regions show the standard error of the mean and highlight that for most of the simulation times, there is a significant difference in the average number of remaining inner planets for each distance.\textbf{ Right:} Histogram of the number of remaining inner planets at 500 Myr for the three set-ups. Starting with four inner planets in all simulations, approx. half of the systems ($\sim$50--60 per cent) keep all of their planets during the simulation time, the remainder of the simulations lose either one or two planets to collisions. We have no simulations where all inner planets collide with each other leaving only one planet.}
     \label{fig:Hist_planet_14_2}
\end{figure*}

\section{Results}\label{results}

There are two main reasons why we would not be able to observe all planets transiting that formed around a star at a later point in time. One, the number of planets at the time of observation is lower than the number of planets when the system formed. Two, the mutual inclinations between the planets are too large for all of these to be within the same observation window. We first analyse the evolution of the average number of planets caused by collisions, after the planetary orbits start to overlap due to changes in eccentricity. This analysis is followed by the evolution of the maximum mutual inclinations. Then, we show how the combination of collisions and mutual inclination differences will affect the number of transiting planets that could be observed from randomly sampled viewing angles, which will be described in the following sections.

\subsection{Number of remaining planets}\label{res_rem_plan}

The number of potentially observable inner planets can be reduced through collisions after orbits start to overlap. We find that none of these close-in planets get ejected in our simulations, which is in agreement with the theory that in the inner region of these systems, collisions happen before ejections \citep[e.g.][]{1972epcf.book.....S,2004ApJ...614..497G}. All collisions are between these planets themselves. We do not see any collisions with either the distant outer Giant, the host or the perturber star. In the left-hand graph in Fig. \ref{fig:Hist_planet_10}, we show the evolution of the average number of inner planets over the simulation time separated for each of the four different planetary architectures with 10 $R_{\rm{MH}}$. The error in the number of planets in the different simulations is calculated using the standard error of the mean. The first collision occurs after $\sim$53 Myr in a simulation with a Giant placed at 5 au. The average number of inner planets drops off continuously for all four-planet system architectures and reaches values of around $\sim$2.2--2.3 planets per system at the end of our simulations (500 Myr) compared to 4 initial inner planets. The simulations without a Giant and one at 5 au show collisions first (55.6 and 53 Myr), followed by the other two architectures (Giant planet at 10 and 20 au) over the next 15 Myr. At 500 Myr, we see that any of the simulations with a Giant have a similar average ($\sim$2.2 planets), whereas the simulations without one have a slightly higher average of remaining planets ($\sim$2.3 planets). However, this difference is not statistically significant, as can be seen from the overlapping standard error regions. On the right-hand graph of Fig. \ref{fig:Hist_planet_10}, we show a histogram of the number of remaining close-in planets at 500 Myr separated by architecture and show the fraction of simulations on the y-axis. We see that $\sim$80 per cent of all initially four inner planet systems reduce to systems with only two inner planets at 500 Myr for the 10 $R_{\rm{MH}}$ initial separation systems. Of these 2-planet systems, $\sim$38 per cent feature one 15 M$_\oplus$ (a result of two collisions) and one 5 M$_\oplus$ planet, whereas the other $\sim$62 per cent feature two 10 M$_\oplus$ planets (two planets with one collision each). We also find a single 20 M$_\oplus$ Super-Earth after three subsequent collisions in a simulation with a Giant planet at 5 au. 

Fig. \ref{fig:Hist_planet_10_2} shows that for the fly-bys with the closest encounter distance of 50 au the planet-planet collisions start the earliest (at $\sim$53 Myr) in the 10 $R_{\rm{MH}}$ simulations, followed by collisions in the 150 au and the 250 au simulations (at $\sim$61 Myr and 78 Myr). None of the systems with a 150 au fly-by retain their initial four inner planets at 500 Myr, whereas some of the 50 au and 250 au keep their initial four inner planet configuration intact until this simulation time. The 50 au and 150 au simulations show a similar drop-off and final value of the average number of inner planets with slightly different starting times. We plot the standard error of the mean which shows that for most of the simulation time, the average number of remaining planets is significantly different for the three different fly-by distances.

Compared to the 10 $R_{\rm{MH}}$ separation simulations, the 14 $R_{\rm{MH}}$ simulations show the first planet-planet collisions later (at $\sim$150 Myr) leading to a higher average number of planets at the end of our simulations. In Fig. \ref{fig:Hist_planet_14} on the right, we see that about half of the systems retain their four planets until 500 Myr. In the left plot of that figure, we see that the simulations with a Giant at 20 au show collisions first, followed a few ten Myr later by the simulations without a Giant planet and the other initial Giant placements. However, the simulations with a Giant planet at 5 and 10 au show a lower average number of remaining inner planets after 500 Myr, having gone through a slightly steeper decline in planet numbers, i.e. more collisions. The shaded regions in this graph represent the standard error of the mean and indicate that there is no significant difference in the average number of remaining planets between the four planet architectures.

When analysing the effect of the encounter distance in Fig. \ref{fig:Hist_planet_14_2} for the 14 $R_{\rm{MH}}$ simulations, we find a much clearer and significant separation in the start of collisions by fly-by distance compared to the 10 $R_{\rm{MH}}$ simulations. Collisions for the 50 au fly-bys start $\sim$90 Myr earlier than those in the 150 au fly-bys, which are ahead of the 250 au fly-bys by another $\sim$40 Myr. The average number of planets at 500 Myr shows clear, significant differences between the fly-by distances with most of the 250 au simulations only going through one or no planet-planet collision. However, the averages in the 50 and 150 au simulations towards the end of the simulation cannot be clearly distinguished from each other due to their overlap in the standard error of the mean. When we look at the histogram on the right of that figure, we find that in the 250 au fly-by simulations only $\sim$10 per cent go through two collisions, making two-planet systems the exception amongst mostly three- and four-planet systems.

\begin{figure*}
    \centering
    \begin{minipage}[t]{1.0\columnwidth}
         \centering
    	\includegraphics[width=1.0\linewidth]{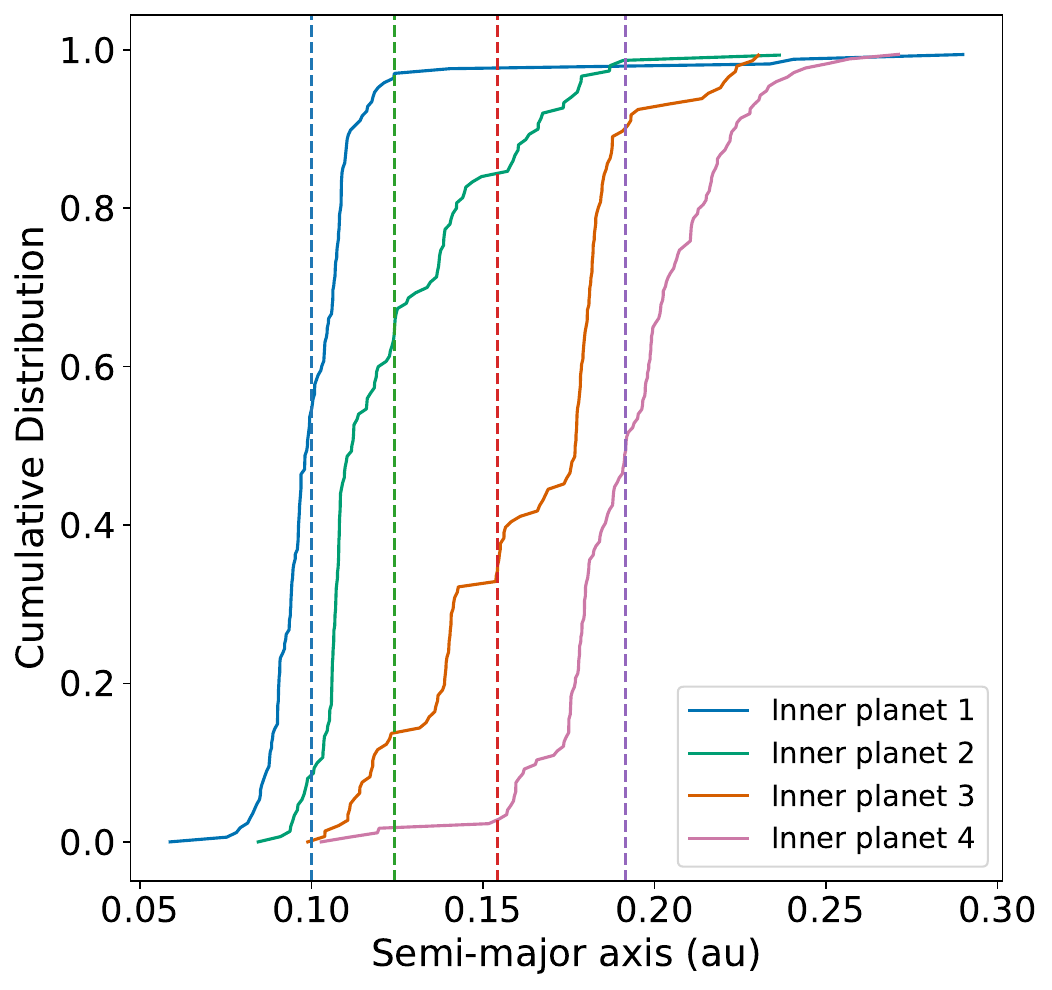}
     \end{minipage}
        \centering
        \vspace{0pt}
    \begin{minipage}[t]{1.0\columnwidth}
    	\includegraphics[width=0.96\linewidth]{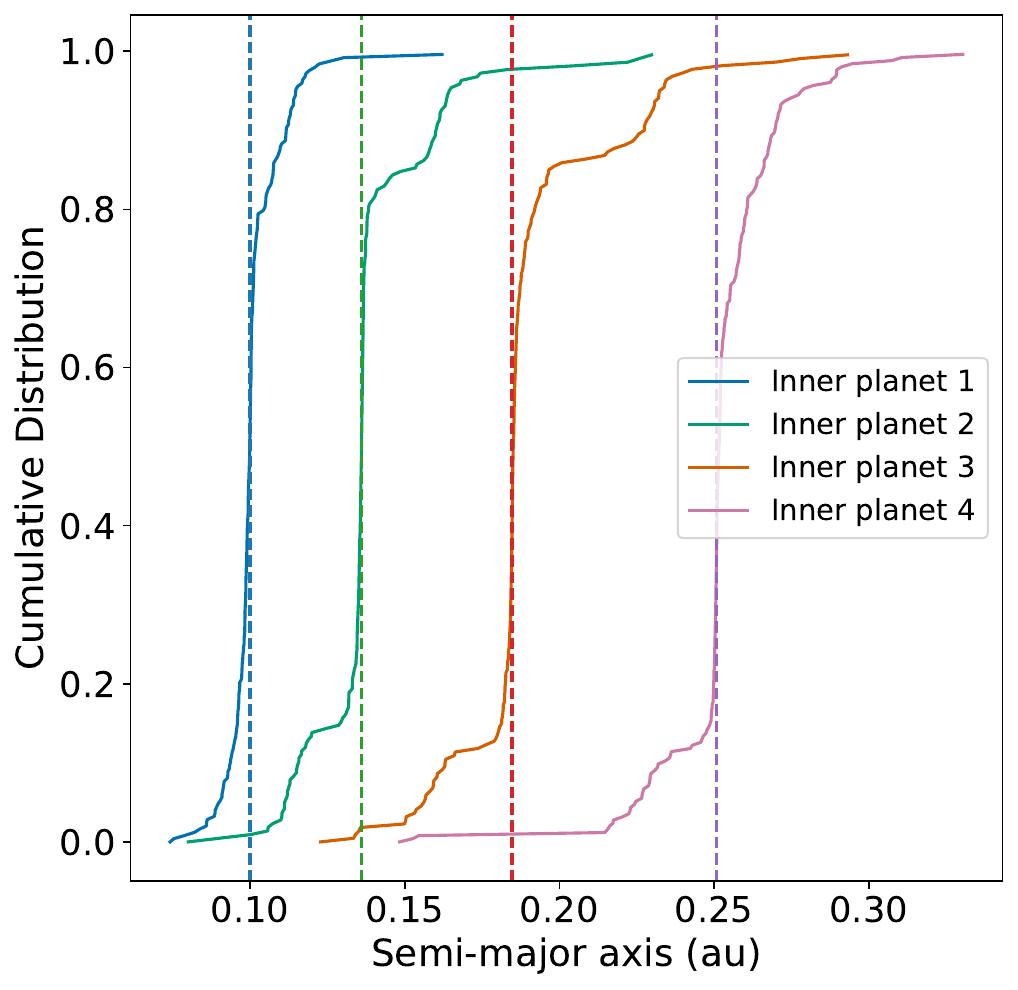}
      \end{minipage}
     \caption{\textbf{Left:} Semi-major axes of remaining planets (after collisions) in all 10 $R_{\rm{MH}}$ simulations at 500 Myr. We plot the original values of the semi-major axes as dashed coloured lines for the 4 inner planets. As shown in Fig. \ref{fig:Hist_planet_10} and \ref{fig:Hist_planet_10_2}, most of the planet systems undergo at least one collision between the planets. These collisions are evidence of highly disrupted systems with overlapping orbits. These changes in the orbits are clearly evident in this cumulative distribution plot. Only a small proportion of the planets are still located at their initial semi-major axis. \textbf{Right:} Semi-major axes of remaining planets (after collisions) in all 14 $R_{\rm{MH}}$ simulations at 500 Myr. We plot the original values of the semi-major axes as dashed coloured lines for the 4 inner planets. As shown in Fig. \ref{fig:Hist_planet_14} and \ref{fig:Hist_planet_14_2}, the majority of the planet systems retain all of their planets in the simulations, showing a lower level of disturbance, which is evident in their semi-major axis values. About 50 per cent of the planets are still located at their original position. These are the planets in simulations without a collision. }
     \label{fig:CDF_semax_10_14}
\end{figure*}

\subsubsection{Orbital excitation of the inner planets}

Comparing the left and right graphs in Fig. \ref{fig:CDF_semax_10_14}, which show the semi-major axes of all remaining inner planets after 500 Myr, we find that the majority of planets in the 14 $R_{\rm{MH}}$ simulations can still be found at their original semi-major axis (right). Whereas virtually all of the planets in the 10 $R_{\rm{MH}}$ simulations (left) have moved to positions either closer or further away from the host star. Based on this orbital evolution of the inner planet system, larger excitation in the orbits of the planets in the initially 10 $R_{\rm{MH}}$ simulations are the consequence. Due to their initially closer spacing, the planets are more easily able to collide with each other. In these plots, we only show the surviving planets and scale them to 1.0, so the number of planets in the right graph of Fig. \ref{fig:CDF_semax_10_14} is much higher than the one on the right.

\begin{figure}
    \centering
    \vspace{0pt}
    \includegraphics[width=1.0\columnwidth]{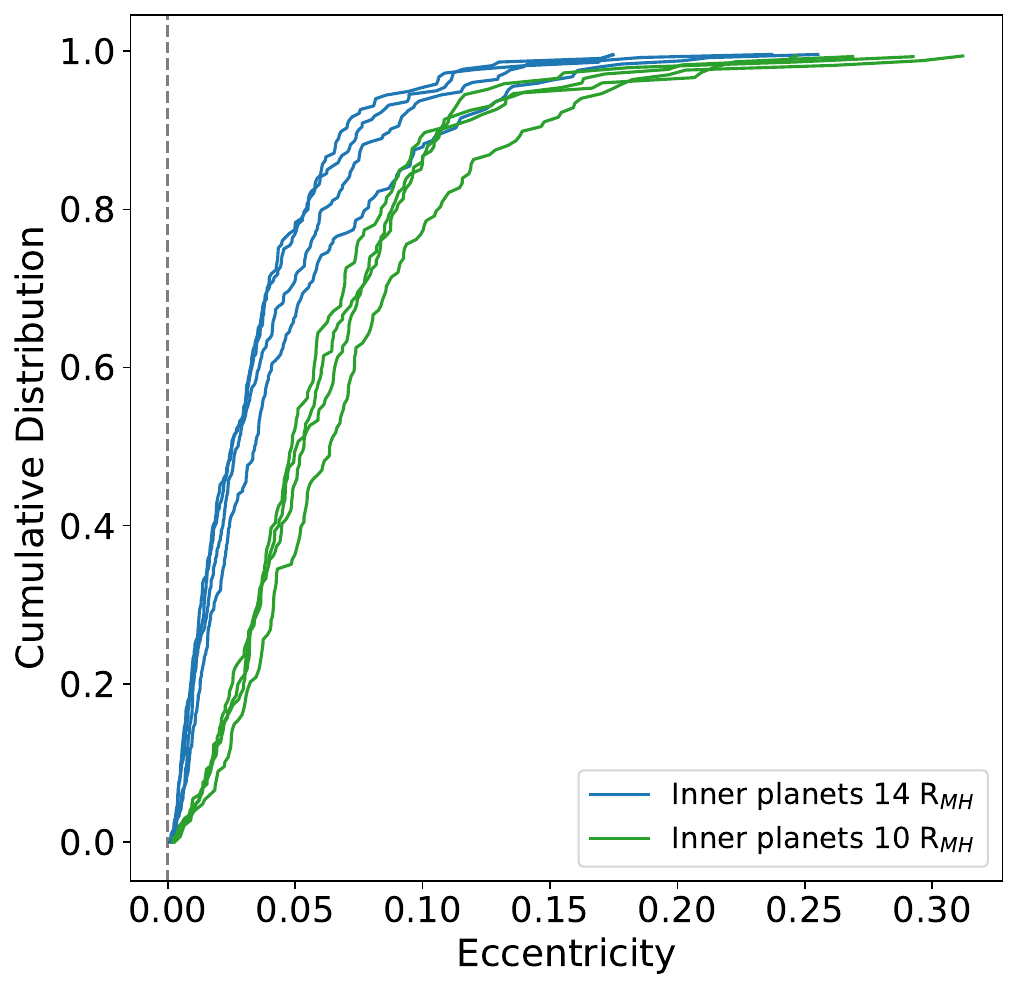}
    \caption{Eccentricity of the remaining inner planets in all 10 and 14 $R_{\rm{MH}}$ simulations at 500 Myr (initial value of zero plotted as black dashed line). Almost none of the planets are still on their initially circular orbit. The planets with 10 $R_{\rm{MH}}$ initial separation ('green') show larger eccentricities than those that were placed initially with 14 $R_{\rm{MH}}$ separation ('blue').}
    \label{fig:CDF_ecc_10_14}
\end{figure}


In Fig. \ref{fig:CDF_ecc_10_14}, we show the eccentricities of the remaining planets after 500 Myr of simulation as a cumulative distribution. We plot both initial separation set-ups together and find that the planet systems with an initial separation of 10 $R_{\rm{MH}}$ show higher eccentricities for all remaining inner planets than for those simulations set up with an initial separation of 14 $R_{\rm{MH}}$. This further illustrates the higher level of excitation that is the result of their initial set-up.

\subsection{Mutual inclination and simulated transits}\label{res_mut_inc}

The probability of a planet transiting its host star on a circular orbit can be calculated using $R_{\ast}/a$ \citep{1984Icar...58..121B}, with $R_{\ast}$ being the stellar radius and $a$ the semi-major axis of the planet's orbit. Based on this relation, we use the approach as shown in \citet{2010ApJ...712.1433B} to calculate the largest angle $\theta$ at which a planet on a circular orbit just transits its host star, i.e. its impact parameter is still below 1:
\begin{equation}\label{Circ_tranit_angle}
\theta = \arcsin \left( \cfrac{R_{\ast}}{a} \right).
\end{equation}
The above shows that the smaller the semi-major axis of the planet in question, the higher its transit probability. To account for non-circular cases that occur when the systems have been disrupted, we modify Eq. \ref{Circ_tranit_angle} to $R_{\ast}/r$, where the semi-major axis $a$ is replaced by the actual distance $r$ between the star and the planet along the line of sight in eccentric orbits \citep[e.g.][]{2010exop.book...15M}:
\begin{equation}\label{Ecc_tranit_angle}
r = \cfrac{a(1-e^2)}{1 + e \cos f},
\end{equation}
Here, $e$ is the eccentricity of the planet's orbit and $f$ is the true anomaly. A planet will transit its host star if its inclination $i$ is located within $\pm \theta \degree$ of the observing direction. By definition, $i$ is expressed in reference to the plane of the sky. For an observer on Earth, a transit will be visible when viewing the star-planet system roughly ``edge-on'', i.e. $(\text{90}\degree - \theta) \leq i \leq (\text{90}\degree + \theta)$ \citep{2010ApJ...712.1433B}. After evaluating the mutual inclination evoluiont, we then use simulated transit observations to determine the number of transiting planets. Instead of using a 90$\degree$ viewing angle, we use a random sample of 1000 different viewing inclinations located within $\pm$90$\degree$ of the highest/lowest planet inclination from our simulations. Even with this favouring of viewing angles located towards the actual transit plane of the planets, for most of the sampled viewing inclinations, no planets will be within the transit zone indicated by $\pm\theta$. 

We then evaluate if our simulated transits could be recovered as shown in several injection/recovery experiments using the Kepler pipeline \citep[e.g.][]{2015ApJ...810...95C,2017AJ....154..109F}. We start by calculating an approximate Signal-to-Noise Ratio (SNR) $m_i$ following Eq. 2 in \citet{2017AJ....154..109F}:
\begin{equation}\label{SNR_injected}
m_i = \left( \cfrac{R_{\rm{P}}}{R_{\ast}} \right)^2 \sqrt{ \cfrac{T_{\rm{obs},i}}{P}}\, \left( \cfrac{1}{\rm{CDPP}_{\rm{dur},i}} \right),
\end{equation}
where $R_{\rm{P}}$ and $P$ are an injected planet's radius and period, and $R_{\ast}$ is as previously defined. $T_{\rm{obs},i}$ is the total observation time of the star and we chose four years for this value, which is the total lifetime of the primary Kepler mission. The final variable in the above equation is the CDPP$_{\rm{dur},i}$, which is the Combined Differential Photometric Precision \citep{2010ApJ...713L..79K} related to the transit duration for each of the planets. The approximate transit durations for our four inner planets are between $\sim$4.5--6 hours. We download the ``Robust RMS CDPP for a searched transit'' for these durations (4.5, 5.0 and 6.0 hours) from the latest data release in the Kepler database \footnote{DR25, https://exoplanetarchive.ipac.caltech.edu/} restricting the stellar radii for the host stars to 0.8--1.2 R$_{\sun}$. We then calculate median values, which are between 130--115 ppm (decreasing precision with increasing transit duration) and use four, evenly spread values in this range for the approximation of the SNRs. 

Fig. 5 in \citet{2017AJ....154..109F} provides an approximation for the fraction of recovered signals depending on the SNR-value. Based on their analysis, $\sim$100 per cent of the injected signals are expected to be recovered if $m_i\gtrsim$\,15, whereas the probability drops to zero for $m_i \lesssim$\,5. We calculate the SNR values for all planets at times between 0 and 500 Myr and compare them to these two critical values. For all inner planets in the 576 simulations at 0 Myr, $m_i$ is larger than 38, which means all of these transit signals would likely be recovered. The signal recovery changes throughout the simulation time, mainly caused by the reduction in individual planetary radii after collisions, as the SNR depends partly on the planet's radius. When calculating if a planet transits, we multiply an injected transit by the recovery fraction calculated using the $\Gamma$ cumulative distribution function given in Eq. 3 in \citet{2017AJ....154..109F}. As shown in there, smaller SNR values result in a not full recovery of the injected transits, which would lead to non-integer values in the recovered transits. For any injected transit that cannot be fully recovered, we therefore randomly sample a number 10 times between 0 and 1 and count a planet transit for this planet system as recovered if this number is smaller than the calculated recovery fraction. If it is larger this transit is not recovered, reducing the total number of transiting planets for this system. For most of our injected transits, the recovery fraction is above 0.95, which leads to most of them being fully counted.

In addition to calculating if a planet will transit when viewed from different observation angles, we also calculate the mutual inclinations between any of our inner planets throughout the simulation time and analyse them based on different characteristics. In particular, we are interested in the maximum mutual inclination in a simulation and use this as a proxy to track the level of excitation in our simulations and how this could then affect their observability. We choose a boundary value of $\sim$4$\degree$. This value is derived by taking the sum of the $\theta$-values of the inner-most (2.65$\degree$) and outer-most close-in planet (1.39$\degree$) on a circular orbit. The idea behind this approach is that two planets with this minimum difference in mutual inclination would not transit together.

\subsubsection{Evolution of maximum mutual inclination}

\begin{figure*}
    \centering
    \vspace{0pt}
    \includegraphics[width=0.9\textwidth]{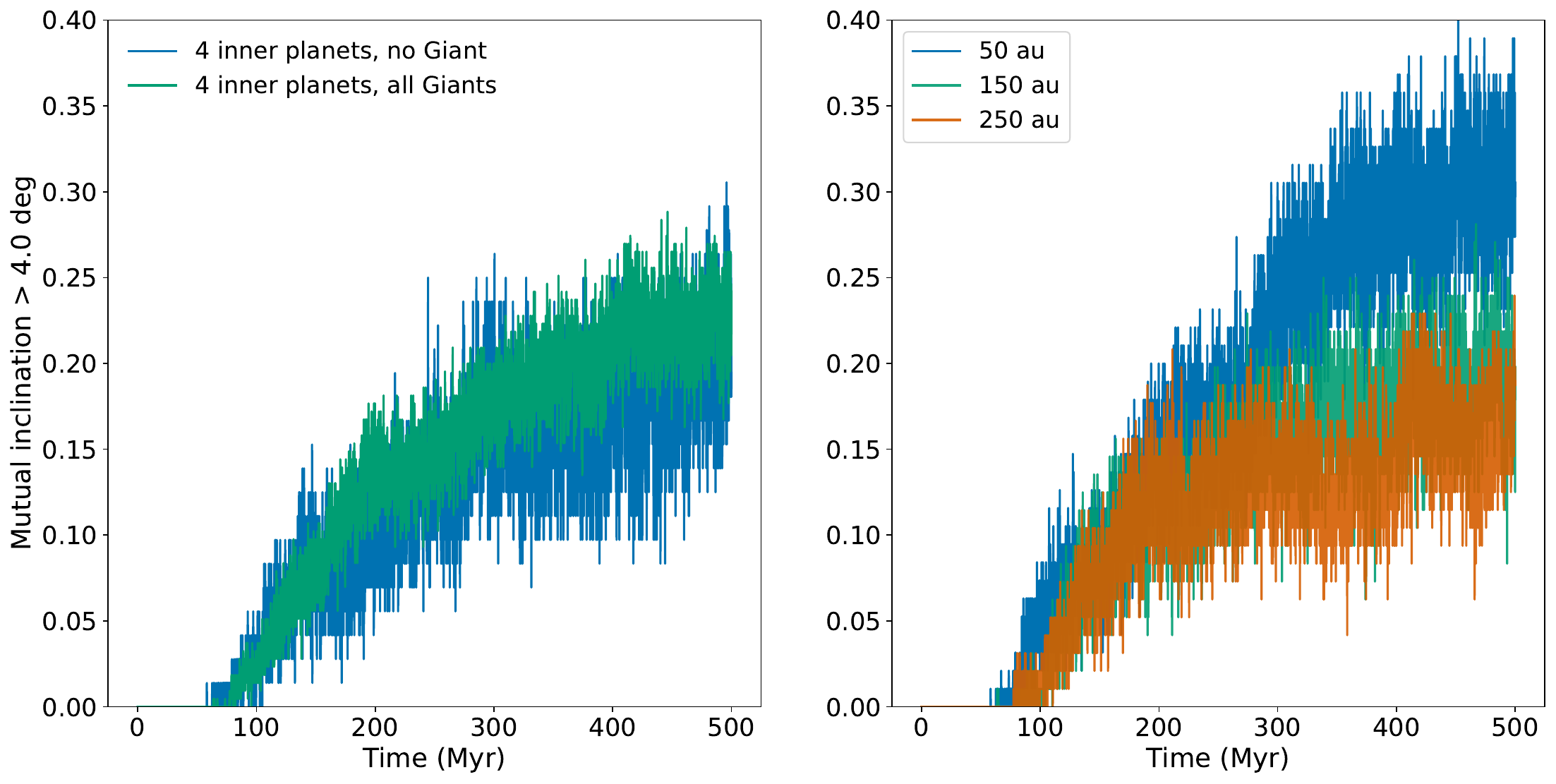}
    \caption{Mutual inclination between the inner planets for 10 $R_{\rm{MH}}$ initial separations. \textbf{Left:} Evolution of the fraction of simulations that show a maximum mutual inclination above 4 $\degree$. The evolution of the four inner planet-only system (blue) is shown separately from the combined ones with an outer Giant case (green). We see that for the first $\sim$50 Myr, none of the simulations have mutual inclinations above the boundary value. Both curves start at about the same time, reaching similar percentages ($\sim$25 per cent) after 500 Myr. The four inner planets-only curve shows a larger range of percentage fluctuations between individual snapshots due to the lower number of simulations (72) compared to the 3 Giant planet set-ups (total of 216). \textbf{Right:} Evolution of the fraction of excited simulations by fly-by distance. The 50 au and 150 au fly-by simulations start to show some disruption as measured by our inclination proxy around the same time, shortly followed by the 250 au simulations. 50 au fly-by simulations reach the highest values ($\sim$35 per cent) after 500 Myr, whereas the 150 and 250 au fly-bys reach lower values $\sim$18 per cent).}
    \label{fig:Mut_Hill10_2plots}
\end{figure*}

\begin{figure*}
        \centering
        \vspace{0pt}
        \includegraphics[width=0.9\textwidth]{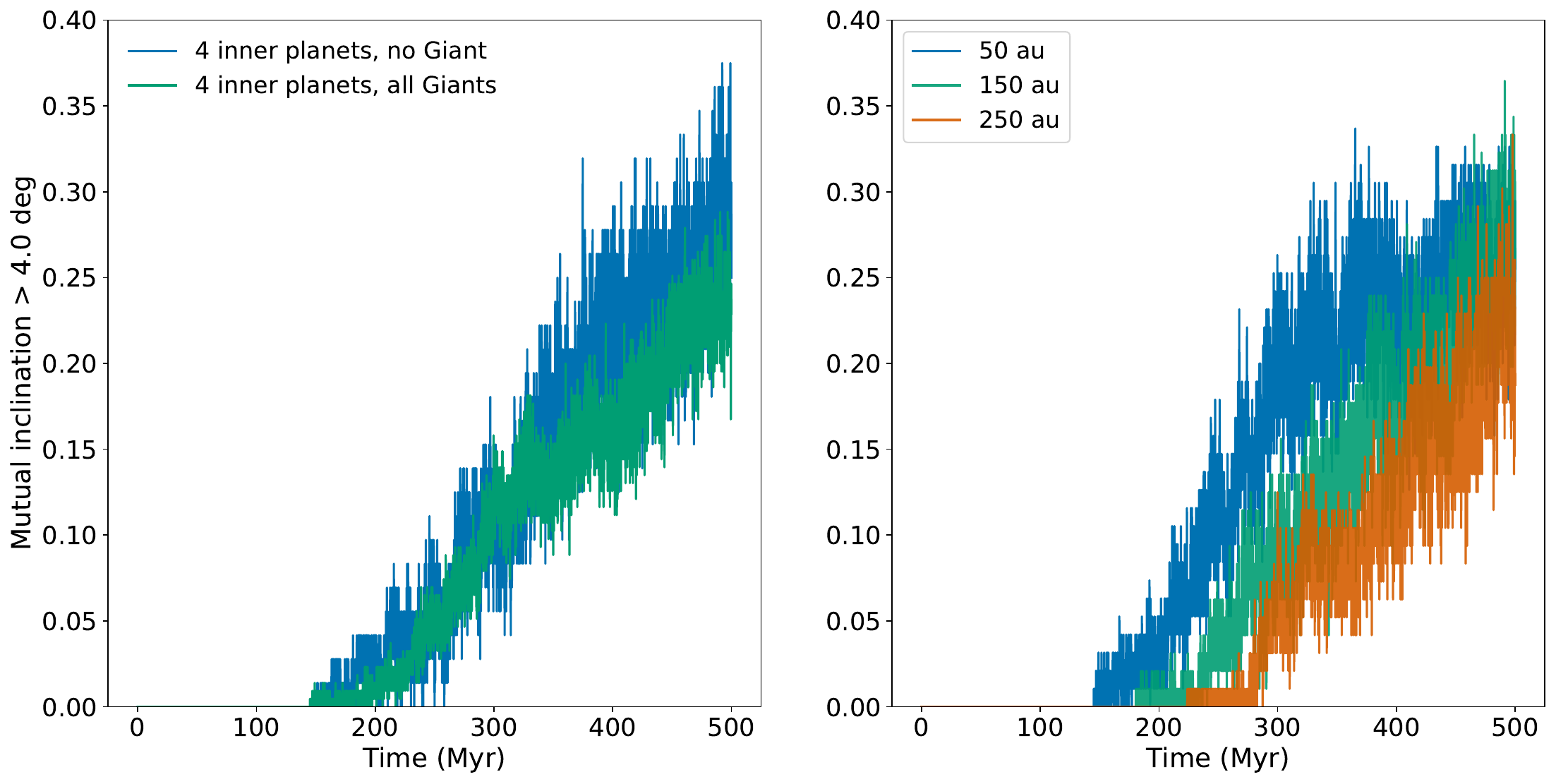}
    \caption{Mutual inclination between the inner planets for 14 $R_{\rm{MH}}$ initial separation. \textbf{Left:} Evolution of the fraction of simulations that show a maximum mutual inclination above 4.0 $\degree$ for any combination of the inner planets. The evolution of the four inner planet-only case (blue) is shown separately from the combined ones with an outer GP case (green). Both curves start to show an increase in disrupted systems around the same time ($\sim$150 Myr). The fraction steadily rises to $\sim$25--30 per cent after 500 Myr. Compared to the 10 $R_{\rm{MH}}$ simulations, the 14 $R_{\rm{MH}}$ ones reach higher values and do not show a flattening of the steepness of the curves. This is likely due to the much lower number of collisions and we expect this development to continue with a longer simulation time. \textbf{Right:} Evolution of the fraction of excited simulations by fly-by distance. There is a clear difference in when the simulations show the first signs of obvious mutual inclination excitation, with closer fly-by simulations disrupting first. The gradients of the three curves appear similar and just offset by their different starting times, reaching similar fractions after 500 Myr.}
    \label{fig:Mut_Hill14_2plots}
\end{figure*}

All our planets start in coplanar orbits with mutual inclinations of 0$\degree$. We track the change in mutual inclinations between the four inner planets over the 500 Myr simulation time. In Fig. \ref{fig:Mut_Hill10_2plots} and \ref{fig:Mut_Hill14_2plots}, we show the fraction of simulations that have a maximum mutual inclination between any two of the inner planets above 4$\degree$ for both initial separation cases.

In the left graph of Fig. \ref{fig:Mut_Hill10_2plots}, we see the maximum mutual inclination values for simulations with initial inner-planet separation of 10 $R_{\rm{MH}}$. We show the evolution of the four-inner-planets-only system separately and combine all four-inner-planets with a Giant at any of the three distances cases. For the first $\sim$50 Myr, none of the simulations have mutual inclinations above the boundary value defined above. Simulations with an outer Giant present start to show higher mutual inclinations first, followed shortly by the first four inner-planet-only simulation a few Myr later at $\sim$58 Myr. In these excited systems, we see collisions following shortly thereafter. The four inner-planet-only plots show a larger spread of fluctuations between individual snapshots due to the lower number of simulations (72) used in the calculation compared to the three Giant planet set-ups (total of 216). While starting to show mutual inclinations departing the coplanar set-up at different times, both reach a value of $\sim$25 per cent after 500 Myr. In the graph on the right in Fig. \ref{fig:Mut_Hill10_2plots}, we show the maximum mutual inclination value depending on the encounter distance over the simulation time. Mutual inclinations above the boundary are reached for more simulations with the lower encounter distance of 50 au, which is apparently able to excite the planetary systems more. The 150 au and 250 au fly-by simulations do not show large differences in the time evolution of the mutual inclinations across our sample.


\begin{figure*}
    \centering
    \begin{minipage}[t]{1.0\columnwidth}
         \centering
    	\includegraphics[width=0.95\linewidth]{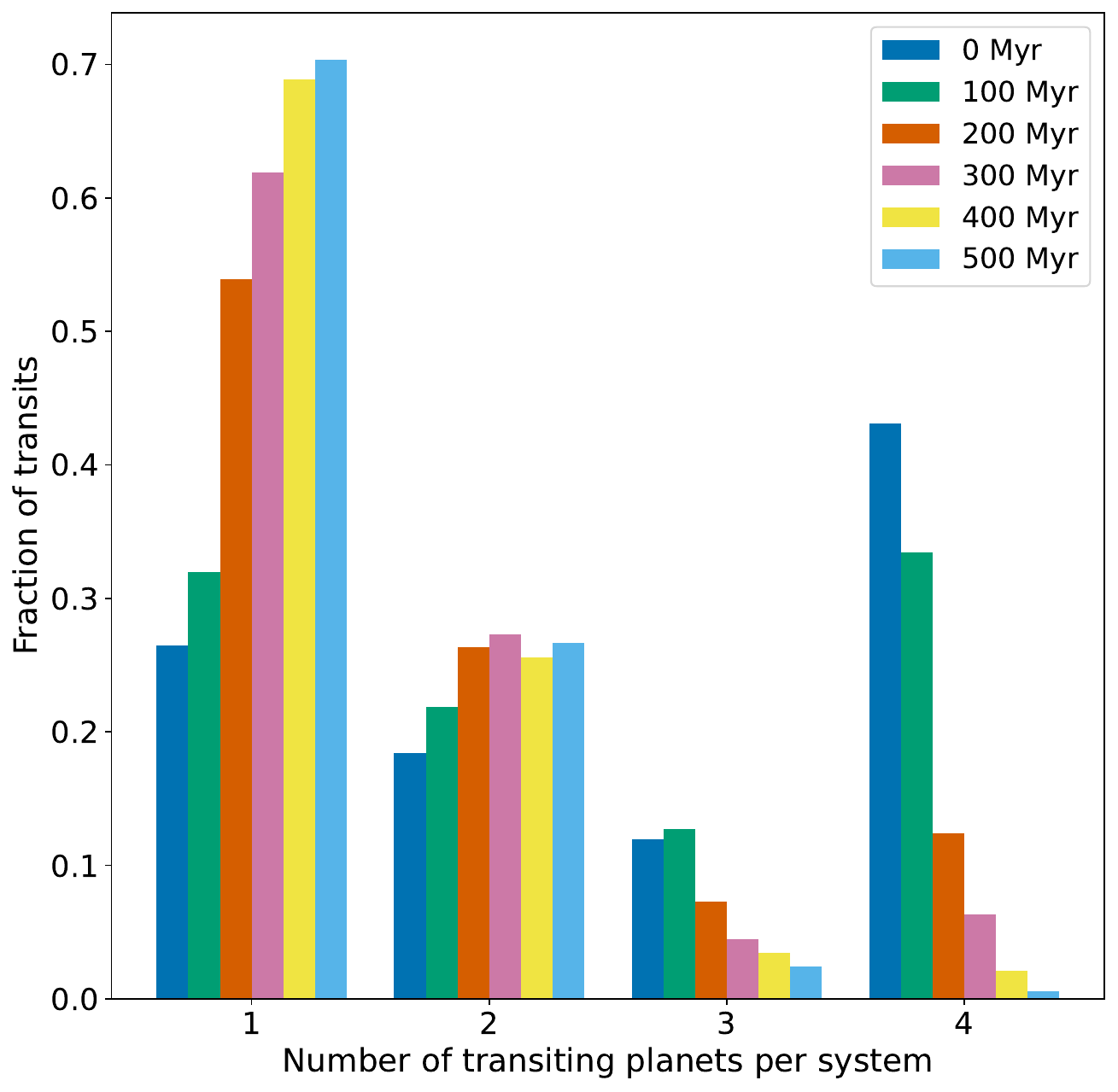}
     \end{minipage}
        \centering
        \vspace{0pt}
    \begin{minipage}[t]{1.0\columnwidth}
    	\includegraphics[width=0.95\linewidth]{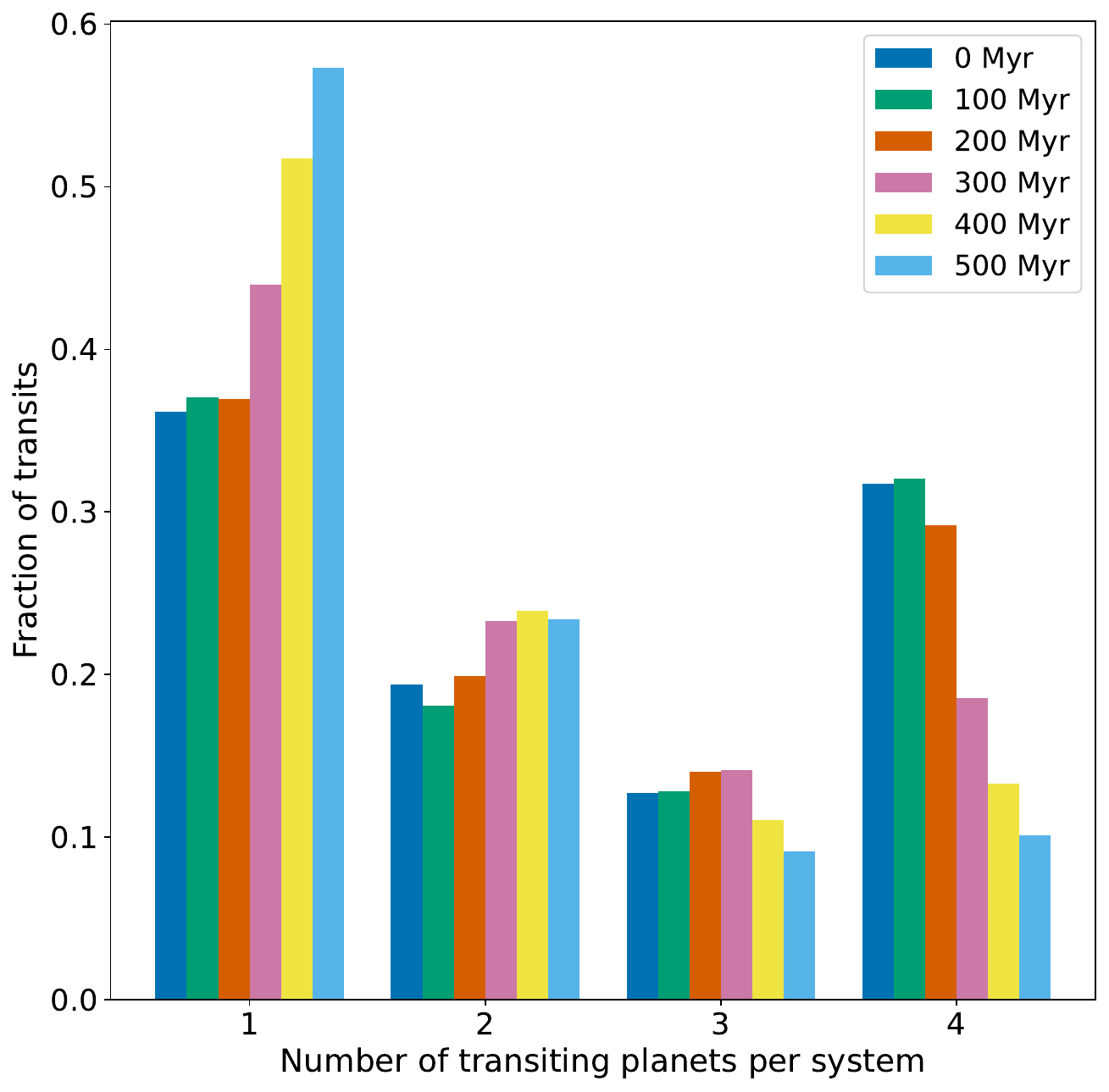}
      \end{minipage}
     \caption{Histogram showing the number of transiting planets from our simulated transit observations over the 500 Myr simulation time. On the \textbf{left}, we see the 10 $R_{\rm{MH}}$ simulations. At 0 Myr, most of the four inner planets transit together as multi-transit systems. The largest change in the transit number occurs between 100-200 Myr, where many simulated systems move from a four-planet transit to a single transit. This is caused by a combination of higher mutual inclinations and a reduction in the number of remaining planets. Towards the end of the simulations at 500 My, virtually no four-transit systems remain, most have evolved into single or two-planet transiting systems. We also see a flattening in the number of new single-transiting systems between 400-500 Myr. On the \textbf{right}, we show the 14 $R_{\rm{MH}}$ simulations. At 0 Myr, slightly more of the four inner planets transit as singles compared to the multi-transit systems due to the larger initial separation, making it easier to only see the inner planet. We see an increase in the number of single-transit systems only after the 200 Myr mark. However, this is not as steep as in the 10 $R_{\rm{MH}}$ simulations in the left graph in Fig. \ref{fig:Hist_transit_all_archi_10_14}. At 500 Myr, there are considerably more single-transit systems than systems with a higher number of transiting planets, despite most of the planetary systems still containing more than one remaining inner planet.}
     \label{fig:Hist_transit_all_archi_10_14}
\end{figure*}

The fraction of simulations showing high maximum mutual inclination in the 14 $R_{\rm{MH}}$ set-up is shown in Fig. \ref{fig:Mut_Hill14_2plots}. The fly-bys happen at the same time as in the 10 $R_{\rm{MH}}$ simulations, but it takes much longer for perturbations to show a measurable effect. The larger initial separations allow these systems to remain coplanar for longer. While the curves start increasing at much later times ($\sim$100 Myr), they reach higher values much quicker than the 10 $R_{\rm{MH}}$ simulations, indicating that more simulations are excited long after the fly-by. The inner planet-only simulations show slightly higher excitation compared to the systems with a Giant planet present and no flattening off as the simulations in Fig. \ref{fig:Mut_Hill10_2plots} did. Regarding the fly-by distance, the closer the fly-by, the sooner we see simulations with higher mutual inclinations.  This is in line with the staggered start of collisions that we have shown earlier in Fig. \ref{fig:Hist_planet_14_2}. The final fraction of excited simulations by fly-by difference is very similar at 500 Myr.

\subsubsection{Disruptive effect of other perturber characteristics}

In addition to the effect of the distance of the fly-by and the planet system architecture, we also evaluate the effect of other perturber characteristics on the maximum mutual inclination between any two of the four inner planets. We find that this proxy can be used to estimate the level of excitation as collisions between planets occur shortly after a simulated planet system passes the 4$\degree$ boundary. For the 10 $R_{\rm{MH}}$ simulations, the first collisions occur at $\sim$53 Myr, which is just one Myr after the first max. mutual inclination simulations passed our 4$\degree$ mark. For the 14 $R_{\rm{MH}}$ simulations, the first collision occurred at $\sim$151 Myr, which is also just a few Myr after passing the same mutual inclination mark. This suggests that our planetary systems remain in an excited, but pre-collision state only for a brief amount of the covered evolution period.

We test three different perturber masses. The simulations with higher-mass perturbers show higher and earlier excitation indicated by an increase in their simulations passing our mutual inclination boundary. For the 10 $R_{\rm{MH}}$ simulations, all three different perturbers disrupt 15--30 per cent of the simulations to show higher mutual inclinations at 500 Myr. In the 14 $R_{\rm{MH}}$, only the 1.0 and 0.5 M$_{\sun}$ perturbers can disrupt the planet systems during our 500 Myr simulation time, whereas the 0.1 M$_{\sun}$ perturber causes very little excitation during the simulation time. We also use two different fly-by velocities, doubling the amount of time any of the perturbers spends in the proximity of the planetary system for the slower velocity fly-by. For both $R_{\rm{MH}}$ sets of simulation, the faster fly-bys lead to an earlier start of the mutual inclinations to \textcolor{changes}{surpass our 4$\degree$ boundary (see Fig. \ref{fig:Mut_inc_vel_evolv} in the appendix)}. However, the fraction of excited systems for the lower velocity fly-bys is offset just by a later start of the disruption, reaching similar values just at later times.

We also change the angle between the perturber and the planetary system orbit. The excitation after 500 Myr as indicated by the maximum mutual inclination in any of our simulations is not correlated with this angle in either of the two $R_{\rm{MH}}$ simulation set-ups. Finally, we analyse if the orbital starting position of the inner planets, either lined up or with $1/4$ of an orbit separation, makes a difference in the amount of excitation. We find that the orbital starting position does not affect the amount of disruption within the planetary systems in either of the two $R_{\rm{MH}}$ simulation sets at all, as would be expected from a secular interaction.

\begin{figure*}
    \centering
    \begin{minipage}[t]{1.0\columnwidth}
         \centering
    	\includegraphics[width=0.95\linewidth]{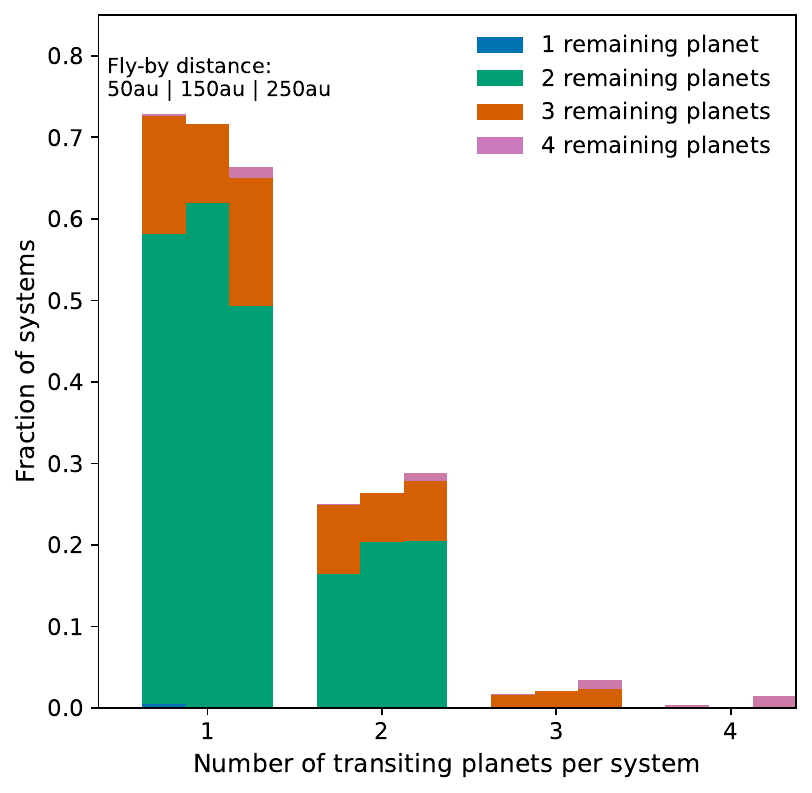}
     \end{minipage}
        \centering
        \vspace{0pt}
    \begin{minipage}[t]{1.0\columnwidth}
    	\includegraphics[width=0.95\linewidth]{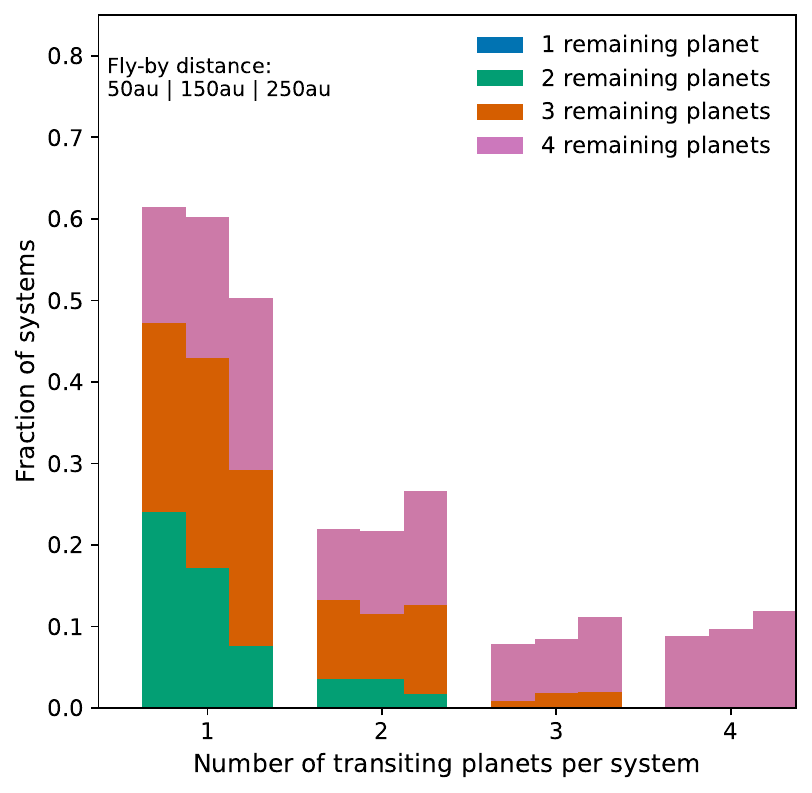}
      \end{minipage}
     \caption{Histogram showing the number of transiting planets for the three different fly-by distances at 500 Myr. On the \textbf{left}, we see the 10 $R_{\rm{MH}}$ simulations. In addition, we also show the number of remaining planets per system for the different transit numbers. As seen in the left graph in Fig. \ref{fig:Hist_transit_all_archi_10_14}, we have virtually no four-transit systems remaining. Those that are, contain some of the few systems that retain their four original inner planets. Most of the simulations now show either single or two-transiting planets at this point in time, regardless of the flyby distance. For the single-transit systems, the 50 and 150 au fly-bys have a higher number of occurrences, compared to the 250 au fly-bys. Both (single and two-transiting planets) systems have a large number of two-planet systems ('green`) as their remaining underlying architecture, with the remainder being made up of three-planet systems ('red`) and a few remaining four-planet systems ('purple`). On the \textbf{right}, we show the 14 $R_{\rm{MH}}$ simulations. We see the highest number of single-transit from simulations with a 50 au fly-by followed by 150 au and then 250 au distances. We end the simulations with $\sim$50 per cent of all simulated systems with four remaining planets at 500 Myr. Despite this, not all of these systems show four planet transits but are distributed across a lower number of transiting planets with higher mutual inclinations. We have no systems with a single remaining planet and most systems with two remaining planets ('green`) show only a single plant transiting. Systems with three remaining planets ('red`) show mainly single or two-transits.}
     \label{fig:Hist_transit_all_archi_10_14_fractions}
\end{figure*}

\subsubsection{Transiting planets}\label{res_transits}

Above, we describe how the mutual inclination between the inner planets and the number of transiting planets are connected. The larger the mutual inclinations between the inner planets, the fewer planets will transit together, but even at low or zero mutual inclination, not all planets will transit together for all viewing angles \citep[e.g.][]{2010arXiv1006.3727R}. In Fig. \ref{fig:Hist_transit_all_archi_10_14}, we show the number of transiting planets at different simulation times (0, 100, 200, 300, 400 and 500 Myr) for the two $R_{\rm{MH}}$ simulations set-ups. We show all architecture combinations of inner planets and Giants together at different times. These histograms have been adjusted to consider the likelihood of a transit signal being successfully recovered in the Kepler pipeline as described above.

Due to their close orbital spacing for the 10 $R_{\rm{MH}}$ systems, all four inner planets will transit for a larger range of viewing angles at 0 Myr (initial set-up), followed by only the inner-most planet, the inner-most two and the inner-most three planets. The difference between the four-planet transit and inner-most-planet-only transit counts is in favour of the single transit for the 14 $R_{\rm{MH}}$ simulations due to the slightly larger separation between them, directly affecting the angle at which they will transit together.

The mutual inclinations in the 10 $R_{\rm{MH}}$ simulations start to increase much earlier than for the 14 $R_{\rm{MH}}$ simulations, which results in the number of simultaneously transiting inner planets decreasing quickly in the case of these initially tighter spaced inner planets. For the initial separation systems with wider spacing, the decrease of transiting planets becomes more noticeable between 200-300 Myr, when more of the systems become excited. This goes along with a decrease in the average number of planets as well as an increase in the number of simulations with higher mutual inclinations (as evident in Fig. \ref{fig:Hist_planet_14} and the left-hand graph in Fig. \ref{fig:Mut_Hill14_2plots}). At 500 Myr, $\sim$70 per cent of the simulations in the 10 $R_{\rm{MH}}$ separation case show only single transits, with a slow-down of two-planet to single transits evolution between 400--500 Myr. The 14 $R_{\rm{MH}}$ simulations show a similar evolution but at a slower pace and a later starting time. The movement towards fewer transiting planets is a mix of increased mutual inclinations and decreased number of planets due to collisions, which is also evident in the earlier figures showing both of these aspects.

We analyse the distribution of the number of remaining planets depending on their transit observability and fly-by distance in Fig. \ref{fig:Hist_transit_all_archi_10_14_fractions} at 500 Myr. Comparing the left and right plots, we immediately notice that the initially closer separation of the inner planets (left) leads to a larger fraction of single-transiting planets at all three fly-by distances than for the initially larger inner planet separation (right). Also, the number of planets within these transiting systems differ, i.e. for the 10 $R_{\rm{MH}}$ systems, most of the single (and double-) transits are caused by two-planet systems, with smaller fractions of three-planet systems. For the 14 $R_{\rm{MH}}$ systems, two-/three-/four-planet systems make up similar proportions of the single and double-transiting systems. We can see in both graphs that there is a drop in the fraction of single-transiting systems for the 250 au fly-bys, which are either counted as part of the double-transits (left) or triple-transits (right).

\begin{figure*}
    \centering
    \begin{minipage}[t]{1.0\columnwidth}
         \centering
    	\includegraphics[width=0.95\linewidth]{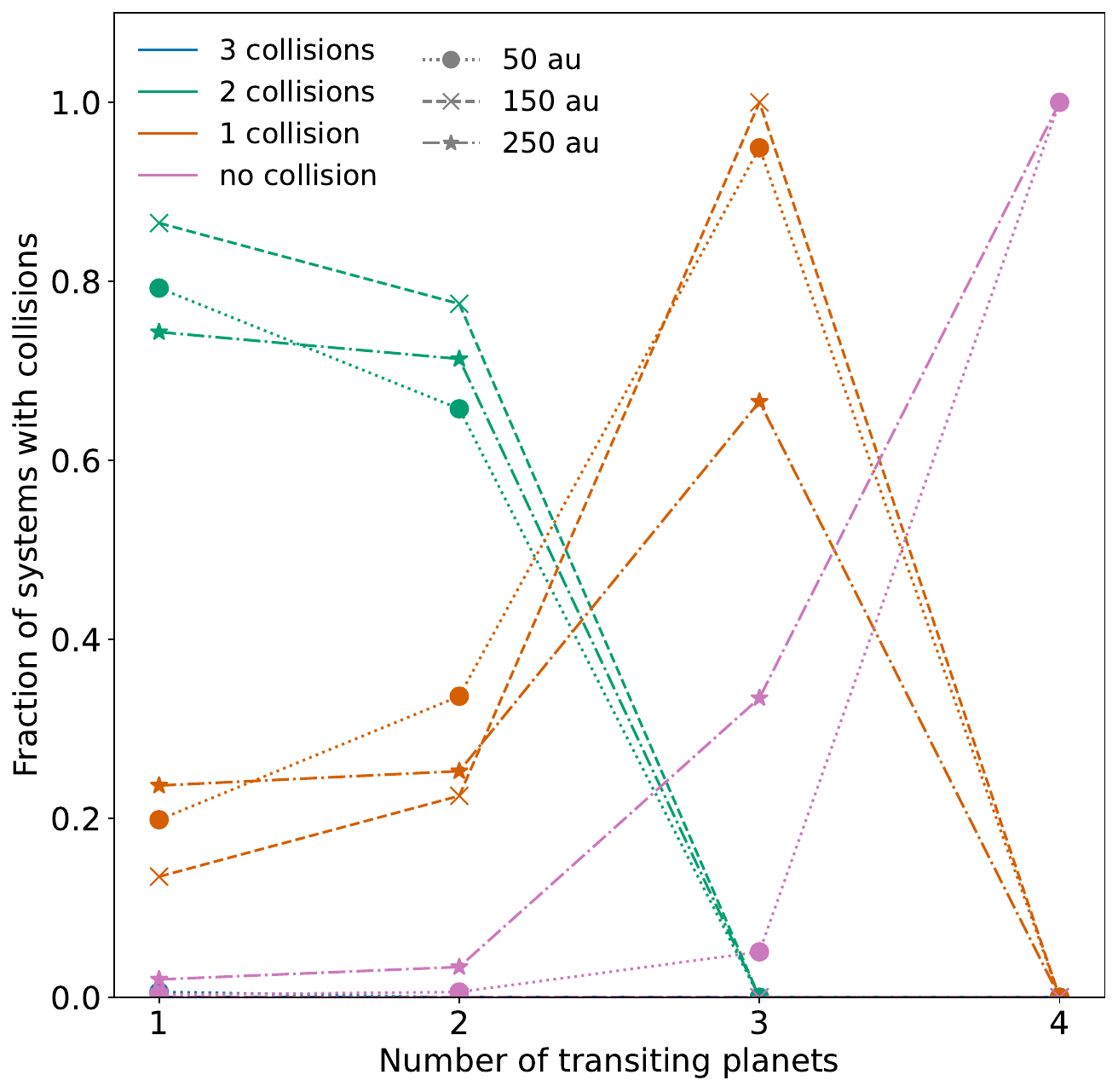}
     \end{minipage}
        \centering
        \vspace{0pt}
    \begin{minipage}[t]{1.0\columnwidth}
    	\includegraphics[width=0.95\linewidth]{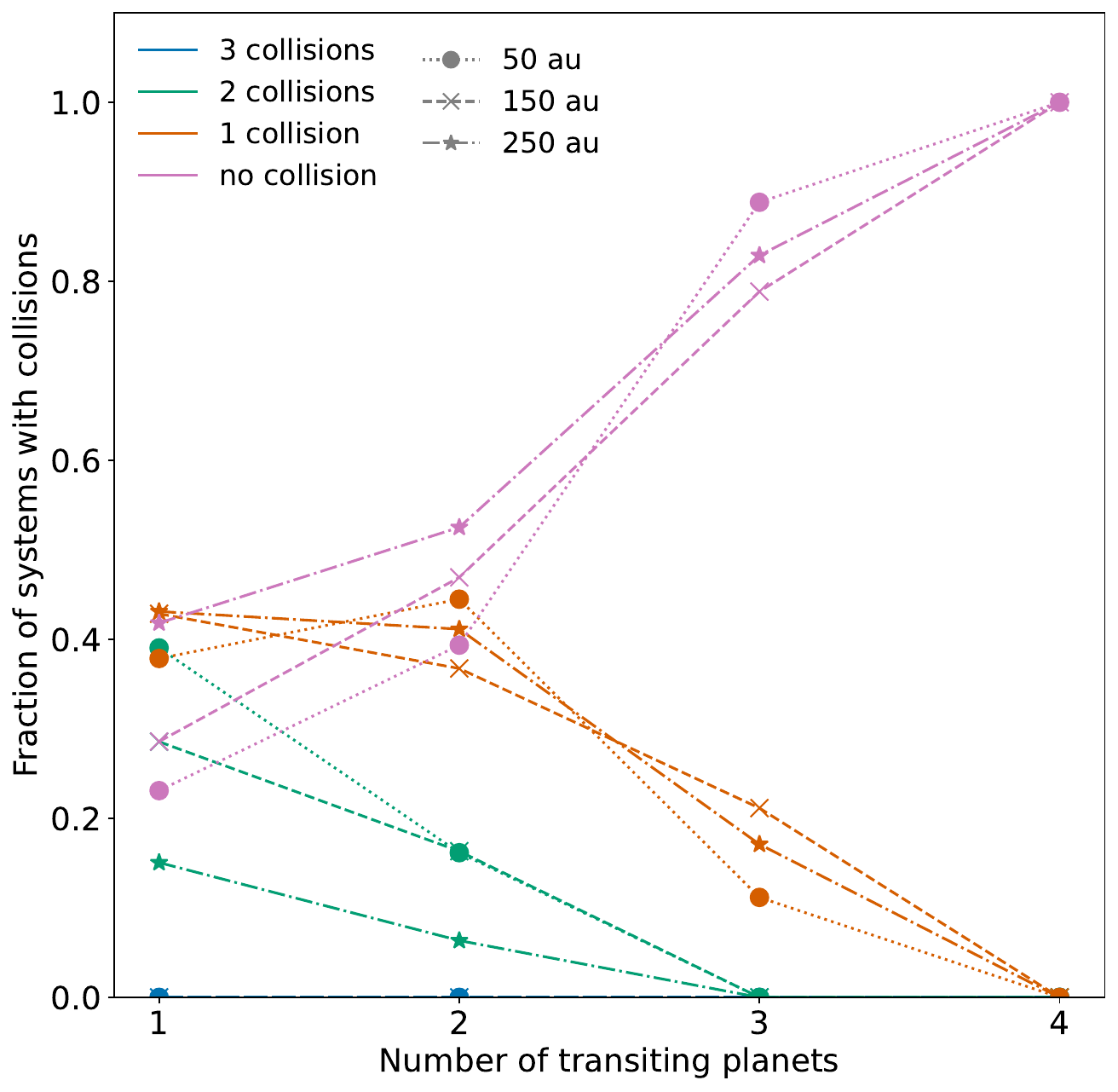}
      \end{minipage}
     \caption{Number of collisions per planetary systems shown for different numbers of transiting planets and fly-by distances at 500 Myr. On the \textbf{left}, we see that all of the four-transit systems are made up of planetary systems that have not experienced any collisions between the inner planets ('purple`), apart from the 150 au simulations which do not show any four-planet transits due to the lack of four-planet systems at 500 Myr. For the three-transit systems, we find that most of the systems have been subject to a fly-by at 50/150 au and have experienced one collision amongst the inner planets ('red`), with a 60:40 split between one and no-collision systems for the 250 au fly-bys. This picture changes for the two-transit systems where 70--80 per cent of all systems have undergone two collisions ('green`), with the remaining transits coming from systems with only one collision. Finally, the single-transit systems have a similar fraction ($\sim$80 per cent) of planetary systems having undergone two collisions for all fly-by distances, with the remainder again made up of single collision systems. On the \textbf{right}, we see that all of the four-transit systems are made up of planetary systems that have not experienced any collisions between the inner planets ('purple`). For the three-transit systems, we find that most of the systems (80--90 per cent) have experienced no collision between the inner planets. For the two-transit systems, we find that $\sim$40--60 per cent of the systems have either had no or one collision, with a negligible number of 2 collisions. For all these different numbers of transiting planets, the fly-by distance has little influence on the transit outcome. Only for the single-transit do we see a change in that 2 collision systems start to become more important. However, this does depend on fly-by distance. For example for the 150 au fly-bys, we see a similar fraction of no collision and two collision systems producing only one transit, whereas for the 50 au fly-bys systems with one or two inner planet collisions show as single of these planets transiting.}
     \label{fig:Hist_transit_all_archi_10_14_collisions}
\end{figure*}

Finally, we have a look at how collisions are distributed amongst the different numbers of transiting planets in Fig. \ref{fig:Hist_transit_all_archi_10_14_collisions}. For the closer separation case (left), we find that single (and double-) transiting systems are predominantly made up of systems having undergone 2 collisions, whereas there is a mix of collision histories for the initially more distant separation case (right). Starting with initially closer inner planets, three transit systems are almost exclusively made up of three planet systems, whereas mutually inclined four-planet systems dominate for the more distant initial separation case.

\section{Discussion}\label{discussion}

In this analysis, we set out to investigate if a typical fly-by that can occur in a young star-forming region can disrupt a system of four close-in planets so that only a single planet transits at any given time. This is done to test if the ``Kepler Dichotomy'' - the apparent excess of single-transiting planets - can, at least partly, be explained as a result of interactions within the birth environment. We do not attempt to fully explain the dichotomy due to several limitations in our approach. One, our simulated planetary systems are still young compared to the planet systems that form the basis for the observed Kepler distribution. Running our simulations, e.g. to 5 Gyr to make them more comparable to the ages of the observed planet systems is not feasible. A second limitation is that we only evolve two example \textcolor{changes}{initially coplanar and circular} planet system architectures under the influence of a perturbing fly-by whereas the Kepler data is the result of a multitude of different initial planet system architectures, which we are unable to observe close to their initial formation state. Finally, \textcolor{changes}{less than 12.5 per cent of the approximately solar-mass initially single stars in our simulations of young star-forming regions have an encounter <300 au after the denser initial core-collapse phase}, thereby leaving a large number of planet systems without this perturbing effect.

The evolution of the number of transits over time in both graphs in Fig. \ref{fig:Hist_transit_all_archi_10_14} shows a clear trend to a lower number of transits for more evolved, older systems. For the 10 $R_{\rm{MH}}$ simulations, most planets are either single or double-transits, which is caused by the reduction in the total number of planets in addition to higher mutual inclinations. These simulations show a slowdown of the increase in the number of single-transit systems from 400 Myr year onwards. Simulating these further in time is required to evaluate if the double-transit systems will move towards single-transit over longer periods of time. This slow-down is also apparent from the flattening of the curve showing the fraction of simulations with mutual inclinations > 4$\degree$, which appear to be caused by the two more distant fly-bys (150 and 250 au).

The 14 $R_{\rm{MH}}$ simulations do not show such a slowdown in the movement toward single transit systems during our simulation time. This is likely due to the later start of the disruption in the planetary systems as the larger initial separations apparently make these systems stable for longer after the fly-by. In these simulations, more than half of the systems still have four planets and it remains to be seen if collisions are as numerous as in the 10 $R_{\rm{MH}}$ simulations. Fig. \ref{fig:Mut_Hill14_2plots} shows that there appears not to be a slowing of the fraction of simulations that show mutual inclinations > 4$\degree$ leading to more and more systems with high enough mutual inclinations to only show single-transits. This suggests that these planetary systems might become more excited over time and move towards fewer transiting planets. 

In the 10 $R_{\rm{MH}}$ simulations, we see a large number of collisions ($\sim$97 per cent of our simulations undergo at least one collision amongst the planets in the inner system) leading to an average number of only $\sim$2.2--2.3 remaining planets. As all of the collisions in our simulations happened between the inner planets, most of these systems now feature either two 10 M$_{\oplus}$ or one even more massive 15 M$_{\oplus}$ Super-Earth. 
Based on our simulations, we suggest that we should find higher-mass SE planets in systems with lower multiplicity and that single-transiting systems would likely have a mix of different planet masses and sizes (either Super-Earths or Sub-Neptunes). The work by \citet{2018AJ....155...48W} ('peas-in-the-pod') found that planets in a multi-planet system are more likely to show similar or correlated sizes, which is similar to the outcome of our simulations, where some of our planetary systems survive with four inner planets of the same size. 

\textcolor{changes}{All of the above are results from initially coplanar and circular planetary systems. We have also run a smaller set of initially slightly eccentric (randomly set for each planet at <0.05) and mutually inclined (randomly set for each of the planets 2--4 at <1$\degree$ mutual inclination compared to the inner-most planet, which has the same initial inclination as in the coplanar simulations to ensure a similar angle between the planet system and the fly-by star). These initial values are motivated by results, e.g. in \citet{2014ApJ...790..146F}, \citet{2015ApJ...808..126V} and \citet{2016PNAS..11311431X}.  The evolution of these simulations (which remained stable in the absence of a flyby) shows similar perturbations leading to increasing mutual inclinations, collisions and the reduction of the number of transiting planets. However, this evolution is accelerated compared to the initially coplanar, circular ones, with an earlier start of the perturbations resulting in a quicker reduction of the number of transiting planets.}

We now compare our findings to the simulations of \citet{2022MNRAS.509.1010R}, who are one of only a few who directly investigated fly-bys as a potential cause for the ``Kepler Dichotomy''. A key difference to our analysis is that these authors only continue to run their simulations for a short period of time after the fly-by, i.e. the time until the perturber is the same distance apart as at the start of the simulation (maximum of 100 times the Giant planet semi-major axis). This was done under the assumption that the largest perturbance is caused by the fly-by itself, even though \citet{2011MNRAS.411..859M} showed in their simulations that it can take tens to hundreds Myr for the perturbance introduced by the fly-by to show a measurable effect. We remove the perturber after 1 Myr, at which point most of them have already reached distances more than $\sim$10$^4$-times that of the Giant's semi-major axis and continue the planet system evolution for another 499 Myr.

\citet{2022MNRAS.509.1010R} simulated close-in inner-planet-systems with one or two Giant planets and found that only the presence of two outer Giants in a fly-by simulation could induce ``sufficient'' inclination misalignment ($\Delta i \gtrsim 0.5\degree$) to cause the inner planets to no longer transit together. Our simulations only include a single Giant planet at three different distances or none at all around a close-in planet system. We show that more than sufficient inclination misalignments (our criteria is $\Delta i \gtrsim 4\degree$) is achieved even in simulations without an outer Giant (as well as those with one) 500 Myr after a fly-by. Our simulations suggest that some level of perturbance appears unavoidable after a fly-by but that it can take a long time before the perturbation causes significant, measurable excitation to the system. 

\textcolor{changes}{In the future it would be interesting to investigate how/if the presence, initial location and mass of the Giant planet impacts the evolution of the inner planet system and how much it is perturbed by the fly-by itself, e.g. testing the hypothesis of an outside-in cascade of inclination perturbance suggested by \citet{2022MNRAS.509.1010R}. This next analysis will also shed light on how the Giant might become inclined or dynamically unstable to cause disruption/perturbation in the inner system \citep[e.g.][]{2017AJ....153...42L,2017MNRAS.467.1531H,2017MNRAS.469..171R,2017MNRAS.468.3000M,2018MNRAS.478..197P,2019MNRAS.482.4146D,2020MNRAS.498.5166P}.}

\citet{2020MNRAS.496.1149L} simulated the fly-by of two star-planet systems with two planets with different configurations. While their simulations are not directly comparable to ours (different number, semi-major axes of planets and mass of their central star), their findings for their close-in Super-Earth systems can be compared to ours. These authors found that their Super-Earths are resistant to the fly-by effect and would require a very close fly-by to destabilise (within a few au). They also suggested that their systems (lower mass than ours) are stable against planet-planet disruption long after the encounter. Our findings agree with the first but disagree with their second conclusion. All our inner planet systems continue to be stable for several tens to hundreds of Myr after the encounter, and none of our fly-by distances are close enough to destabilise them directly, even though the planets are placed close to the stability limit. However, we find that many of our IP systems become unstable enough over time to cause planet-planet collisions, but we see no inner planet ejections.

\section{Conclusions}\label{conclusion}

In this paper, we use $N$-body simulations to investigate the effect of close encounters that can occur during the early dynamical evolution of a typical young star-forming region on planetary systems containing close-in planets (with and without a distant Giant planet) around a solar-mass host star.

\renewcommand{\labelenumi}{(\roman{enumi})}
\begin{enumerate}
  \item A single fly-by at <300 au can excite the orbits of close-in inner planets in such a way that initially, coplanar and circular systems evolve into systems with higher eccentricities and mutual inclinations. These orbital changes can lead to collisions between the inner planets. Both effects then result in a reduction in the number of transiting planets over time.
  \item We find that many of the simulated planet systems have long periods of apparent stability before they become unstable over a very short period of time, ending in collisions and/or with high mutual inclinations. At the end of our simulations (500 Myr), we see a considerable amount of single-transiting systems but do not reproduce the Kepler transit distribution. We did not set out to re-produce this due to our choice of only two basic planetary architectures; however, we show that fly-bys can contribute to the production of single-transit systems by themselves. We also find that a longer simulation time is required to evaluate the evolution, especially of the systems with higher initial orbital separation.
  \item The distances of the fly-bys result in differences in the maximum mutual inclinations for the initially closer inner planet separations but not for the more distant set-up. There is a time-lag in when the systems start to get excited but the result after 500 Myr is roughly similar. The fly-by distance difference has a small effect on the number of collisions in the systems, but this difference is small and not statistically significant in all cases. Other perturber characteristics have similar delay effects to when the inner planet systems get excited, i.e. velocity or perturber mass. But neither the orbital starting position nor the angle between the orbital plane of the inner planets and the perturber produces any difference in the resulting maximum mutual inclinations.
  \item In addition to the initial inner planet separation, we also test four different architectures, where three feature a Giant planet at different locations. This difference does not affect the behaviour of the inner planet systems with regard to collisions or mutual inclinations, as we find no statistical difference between the architecture setups.
\end{enumerate}

\section*{Acknowledgements}

This project has received funding from the European Research Council (ERC) under the European Union’s Horizon 2020 Framework Programme (grant agreement no. 853022, PEVAP). JEO is supported by a Royal Society University Research Fellowship.

Simulations in this paper made use of the REBOUND N-body code \citep{rebound}. The simulations were integrated using the hybrid symplectic MERCURIUS integrator \citep{reboundmercurius} and IAS15, a 15th order Gauss-Radau integrator \citep{reboundias15}. The SimulationArchive format was used to store fully reproducible simulation data \citep{reboundsa}. 

This work was performed using the DiRAC Data Intensive service at Leicester, operated by the University of Leicester IT Services, which forms part of the STFC DiRAC HPC Facility (www.dirac.ac.uk). The equipment was funded by BEIS capital funding via STFC capital grants ST/K000373/1 and ST/R002363/1 and STFC DiRAC Operations grant ST/R001014/1. DiRAC is part of the National e-Infrastructure. \textcolor{changes}{This work was performed using resources provided by the Cambridge Service for Data Driven Discovery (CSD3) operated by the University of Cambridge Research Computing Service (www.csd3.cam.ac.uk), provided by Dell EMC and Intel using Tier-2 funding from the Engineering and Physical Sciences Research Council (capital grant EP/T022159/1), and DiRAC funding from the Science and Technology Facilities Council (www.dirac.ac.uk).}

This research has made use of the NASA Exoplanet Archive, which is operated by the California Institute of Technology, under contract with the National Aeronautics and Space Administration under the Exoplanet Exploration Program.

\section*{Data Availability}

The data underlying this article will be shared on reasonable request
to the corresponding author.



\bibliographystyle{mnras}
\bibliography{Main_document} 

\begin{thebibliography}{}
\makeatletter
\relax
\def\mn@urlcharsother{\let\do\@makeother \do\$\do\&\do\#\do\^\do\_\do\%\do\~}
\def\mn@doi{\begingroup\mn@urlcharsother \@ifnextchar [ {\mn@doi@} {\mn@doi@[]}}
\def\mn@doi@[#1]#2{\def\@tempa{#1}\ifx\@tempa\@empty \href {http://dx.doi.org/#2} {doi:#2}\else \href {http://dx.doi.org/#2} {#1}\fi \endgroup}
\def\mn@eprint#1#2{\mn@eprint@#1:#2::\@nil}
\def\mn@eprint@arXiv#1{\href {http://arxiv.org/abs/#1} {{\tt arXiv:#1}}}
\def\mn@eprint@dblp#1{\href {http://dblp.uni-trier.de/rec/bibtex/#1.xml} {dblp:#1}}
\def\mn@eprint@#1:#2:#3:#4\@nil{\def\@tempa {#1}\def\@tempb {#2}\def\@tempc {#3}\ifx \@tempc \@empty \let \@tempc \@tempb \let \@tempb \@tempa \fi \ifx \@tempb \@empty \def\@tempb {arXiv}\fi \@ifundefined {mn@eprint@\@tempb}{\@tempb:\@tempc}{\expandafter \expandafter \csname mn@eprint@\@tempb\endcsname \expandafter{\@tempc}}}

\bibitem[\protect\citeauthoryear{{Allison} \& {Goodwin}}{{Allison} \& {Goodwin}}{2011}]{RN38}
{Allison} R.~J.,  {Goodwin} S.~P.,  2011, \mn@doi [\mnras] {10.1111/j.1365-2966.2011.18849.x}, 415, 1967

\bibitem[\protect\citeauthoryear{{Allison}, {Goodwin}, {Parker}, {Portegies Zwart}  \& {de Grijs}}{{Allison} et~al.}{2010}]{RN4}
{Allison} R.~J.,  {Goodwin} S.~P.,  {Parker} R.~J.,  {Portegies Zwart} S.~F.,   {de Grijs} R.,  2010, \mn@doi [\mnras] {10.1111/j.1365-2966.2010.16939.x}, 407, 1098

\bibitem[\protect\citeauthoryear{{Alves}, {Cleeves}, {Girart}, {Zhu}, {Franco}, {Zurlo}  \& {Caselli}}{{Alves} et~al.}{2020}]{2020ApJ...904L...6A}
{Alves} F.~O.,  {Cleeves} L.~I.,  {Girart} J.~M.,  {Zhu} Z.,  {Franco} G. A.~P.,  {Zurlo} A.,   {Caselli} P.,  2020, \mn@doi [\apjl] {10.3847/2041-8213/abc550}, 904, L6

\bibitem[\protect\citeauthoryear{{Beatty} \& {Seager}}{{Beatty} \& {Seager}}{2010}]{2010ApJ...712.1433B}
{Beatty} T.~G.,  {Seager} S.,  2010, \mn@doi [\apj] {10.1088/0004-637X/712/2/1433}, 712, 1433

\bibitem[\protect\citeauthoryear{{Biersteker} \& {Schlichting}}{{Biersteker} \& {Schlichting}}{2019}]{2019MNRAS.485.4454B}
{Biersteker} J.~B.,  {Schlichting} H.~E.,  2019, \mn@doi [\mnras] {10.1093/mnras/stz738}, 485, 4454

\bibitem[\protect\citeauthoryear{{Borucki} \& {Summers}}{{Borucki} \& {Summers}}{1984}]{1984Icar...58..121B}
{Borucki} W.~J.,  {Summers} A.~L.,  1984, \mn@doi [\icarus] {10.1016/0019-1035(84)90102-7}, 58, 121

\bibitem[\protect\citeauthoryear{{Breslau} \& {Pfalzner}}{{Breslau} \& {Pfalzner}}{2019}]{2019A&A...621A.101B}
{Breslau} A.,  {Pfalzner} S.,  2019, \mn@doi [\aap] {10.1051/0004-6361/201833729}, 621, A101

\bibitem[\protect\citeauthoryear{{Bressert} et~al.,}{{Bressert} et~al.}{2010}]{RN59}
{Bressert} E.,  et~al., 2010, \mn@doi [\mnras] {10.1111/j.1745-3933.2010.00946.x}, 409, L54

\bibitem[\protect\citeauthoryear{{Brown} \& {Rein}}{{Brown} \& {Rein}}{2022}]{2022MNRAS.515.5942B}
{Brown} G.,  {Rein} H.,  2022, \mn@doi [\mnras] {10.1093/mnras/stac1763}, 515, 5942

\bibitem[\protect\citeauthoryear{{Bryan} et~al.,}{{Bryan} et~al.}{2016}]{2016ApJ...821...89B}
{Bryan} M.~L.,  et~al., 2016, \mn@doi [\apj] {10.3847/0004-637X/821/2/89}, 821, 89

\bibitem[\protect\citeauthoryear{{Bryan}, {Knutson}, {Lee}, {Fulton}, {Batygin}, {Ngo}  \& {Meshkat}}{{Bryan} et~al.}{2019}]{2019AJ....157...52B}
{Bryan} M.~L.,  {Knutson} H.~A.,  {Lee} E.~J.,  {Fulton} B.~J.,  {Batygin} K.,  {Ngo} H.,   {Meshkat} T.,  2019, \mn@doi [\aj] {10.3847/1538-3881/aaf57f}, 157, 52

\bibitem[\protect\citeauthoryear{{Cai}, {Kouwenhoven}, {Portegies Zwart}  \& {Spurzem}}{{Cai} et~al.}{2017}]{2017MNRAS.470.4337C}
{Cai} M.~X.,  {Kouwenhoven} M.~B.~N.,  {Portegies Zwart} S.~F.,   {Spurzem} R.,  2017, \mn@doi [\mnras] {10.1093/mnras/stx1464}, 470, 4337

\bibitem[\protect\citeauthoryear{{Cai}, {Portegies Zwart}  \& {van Elteren}}{{Cai} et~al.}{2018}]{2018MNRAS.474.5114C}
{Cai} M.~X.,  {Portegies Zwart} S.,   {van Elteren} A.,  2018, \mn@doi [\mnras] {10.1093/mnras/stx3064}, 474, 5114

\bibitem[\protect\citeauthoryear{{Cai}, {Portegies Zwart}, {Kouwenhoven}  \& {Spurzem}}{{Cai} et~al.}{2019}]{2019MNRAS.489.4311C}
{Cai} M.~X.,  {Portegies Zwart} S.,  {Kouwenhoven} M.~B.~N.,   {Spurzem} R.,  2019, \mn@doi [\mnras] {10.1093/mnras/stz2467}, 489, 4311

\bibitem[\protect\citeauthoryear{{Chabrier}}{{Chabrier}}{2005}]{RN200}
{Chabrier} G.,  2005, in {Corbelli} E.,  {Palla} F.,   {Zinnecker} H.,  eds,  Astrophysics and Space Science Library Vol. 327, The Initial Mass Function 50 Years Later. p.~41, \mn@doi{10.1007/978-1-4020-3407-7_5}

\bibitem[\protect\citeauthoryear{{Christiansen} et~al.,}{{Christiansen} et~al.}{2015}]{2015ApJ...810...95C}
{Christiansen} J.~L.,  et~al., 2015, \mn@doi [\apj] {10.1088/0004-637X/810/2/95}, 810, 95

\bibitem[\protect\citeauthoryear{{Cumming}, {Butler}, {Marcy}, {Vogt}, {Wright}  \& {Fischer}}{{Cumming} et~al.}{2008}]{2008PASP..120..531C}
{Cumming} A.,  {Butler} R.~P.,  {Marcy} G.~W.,  {Vogt} S.~S.,  {Wright} J.~T.,   {Fischer} D.~A.,  2008, \mn@doi [\pasp] {10.1086/588487}, 120, 531

\bibitem[\protect\citeauthoryear{{Daffern-Powell}, {Parker}  \& {Quanz}}{{Daffern-Powell} et~al.}{2022}]{2022MNRAS.514..920D}
{Daffern-Powell} E.~C.,  {Parker} R.~J.,   {Quanz} S.~P.,  2022, \mn@doi [\mnras] {10.1093/mnras/stac1392}, 514, 920

\bibitem[\protect\citeauthoryear{{Demory} \& {Seager}}{{Demory} \& {Seager}}{2011}]{2011ApJS..197...12D}
{Demory} B.-O.,  {Seager} S.,  2011, \mn@doi [\apjs] {10.1088/0067-0049/197/1/12}, 197, 12

\bibitem[\protect\citeauthoryear{{Denham}, {Naoz}, {Hoang}, {Stephan}  \& {Farr}}{{Denham} et~al.}{2019}]{2019MNRAS.482.4146D}
{Denham} P.,  {Naoz} S.,  {Hoang} B.-M.,  {Stephan} A.~P.,   {Farr} W.~M.,  2019, \mn@doi [\mnras] {10.1093/mnras/sty2830}, 482, 4146

\bibitem[\protect\citeauthoryear{{Dong}, {Zhu}  \& {Whitney}}{{Dong} et~al.}{2015}]{2015ApJ...809...93D}
{Dong} R.,  {Zhu} Z.,   {Whitney} B.,  2015, \mn@doi [\apj] {10.1088/0004-637X/809/1/93}, 809, 93

\bibitem[\protect\citeauthoryear{{Fabrycky} et~al.,}{{Fabrycky} et~al.}{2014}]{2014ApJ...790..146F}
{Fabrycky} D.~C.,  et~al., 2014, \mn@doi [\apj] {10.1088/0004-637X/790/2/146}, 790, 146

\bibitem[\protect\citeauthoryear{{Flammini Dotti}, {Kouwenhoven}, {Cai}  \& {Spurzem}}{{Flammini Dotti} et~al.}{2019}]{2019MNRAS.489.2280F}
{Flammini Dotti} F.,  {Kouwenhoven} M.~B.~N.,  {Cai} M.~X.,   {Spurzem} R.,  2019, \mn@doi [\mnras] {10.1093/mnras/stz2346}, 489, 2280

\bibitem[\protect\citeauthoryear{{Flammini Dotti}, {Capuzzo-Dolcetta}  \& {Kouwenhoven}}{{Flammini Dotti} et~al.}{2023}]{2023MNRAS.tmp.2714F}
{Flammini Dotti} F.,  {Capuzzo-Dolcetta} R.,   {Kouwenhoven} M.~B.~N.,  2023, \mn@doi [\mnras] {10.1093/mnras/stad2819}

\bibitem[\protect\citeauthoryear{{Fortney}, {Marley}  \& {Barnes}}{{Fortney} et~al.}{2007a}]{2007ApJ...659.1661F}
{Fortney} J.~J.,  {Marley} M.~S.,   {Barnes} J.~W.,  2007a, \mn@doi [\apj] {10.1086/512120}, 659, 1661

\bibitem[\protect\citeauthoryear{{Fortney}, {Marley}  \& {Barnes}}{{Fortney} et~al.}{2007b}]{2007ApJ...668.1267F}
{Fortney} J.~J.,  {Marley} M.~S.,   {Barnes} J.~W.,  2007b, \mn@doi [\apj] {10.1086/521435}, 668, 1267

\bibitem[\protect\citeauthoryear{{Fulton} et~al.,}{{Fulton} et~al.}{2017}]{2017AJ....154..109F}
{Fulton} B.~J.,  et~al., 2017, \mn@doi [\aj] {10.3847/1538-3881/aa80eb}, 154, 109

\bibitem[\protect\citeauthoryear{{Fulton} et~al.,}{{Fulton} et~al.}{2021}]{2021ApJS..255...14F}
{Fulton} B.~J.,  et~al., 2021, \mn@doi [\apjs] {10.3847/1538-4365/abfcc1}, 255, 14

\bibitem[\protect\citeauthoryear{{Goldreich}, {Lithwick}  \& {Sari}}{{Goldreich} et~al.}{2004}]{2004ApJ...614..497G}
{Goldreich} P.,  {Lithwick} Y.,   {Sari} R.,  2004, \mn@doi [\apj] {10.1086/423612}, 614, 497

\bibitem[\protect\citeauthoryear{{Goodwin} \& {Whitworth}}{{Goodwin} \& {Whitworth}}{2004}]{RN14}
{Goodwin} S.~P.,  {Whitworth} A.~P.,  2004, \mn@doi [\aap] {10.1051/0004-6361:20031529}, 413, 929

\bibitem[\protect\citeauthoryear{{Hansen}}{{Hansen}}{2017}]{2017MNRAS.467.1531H}
{Hansen} B. M.~S.,  2017, \mn@doi [\mnras] {10.1093/mnras/stx182}, 467, 1531

\bibitem[\protect\citeauthoryear{{Hao}, {Kouwenhoven}  \& {Spurzem}}{{Hao} et~al.}{2013}]{2013MNRAS.433..867H}
{Hao} W.,  {Kouwenhoven} M.~B.~N.,   {Spurzem} R.,  2013, \mn@doi [\mnras] {10.1093/mnras/stt771}, 433, 867

\bibitem[\protect\citeauthoryear{{He}, {Ford}  \& {Ragozzine}}{{He} et~al.}{2019}]{2019MNRAS.490.4575H}
{He} M.~Y.,  {Ford} E.~B.,   {Ragozzine} D.,  2019, \mn@doi [\mnras] {10.1093/mnras/stz2869}, 490, 4575

\bibitem[\protect\citeauthoryear{{He}, {Ford}, {Ragozzine}  \& {Carrera}}{{He} et~al.}{2020}]{2020AJ....160..276H}
{He} M.~Y.,  {Ford} E.~B.,  {Ragozzine} D.,   {Carrera} D.,  2020, \mn@doi [\aj] {10.3847/1538-3881/abba18}, 160, 276

\bibitem[\protect\citeauthoryear{{Izidoro}, {Ogihara}, {Raymond}, {Morbidelli}, {Pierens}, {Bitsch}, {Cossou}  \& {Hersant}}{{Izidoro} et~al.}{2017}]{2017MNRAS.470.1750I}
{Izidoro} A.,  {Ogihara} M.,  {Raymond} S.~N.,  {Morbidelli} A.,  {Pierens} A.,  {Bitsch} B.,  {Cossou} C.,   {Hersant} F.,  2017, \mn@doi [\mnras] {10.1093/mnras/stx1232}, 470, 1750

\bibitem[\protect\citeauthoryear{{Izidoro}, {Bitsch}, {Raymond}, {Johansen}, {Morbidelli}, {Lambrechts}  \& {Jacobson}}{{Izidoro} et~al.}{2021}]{2021A&A...650A.152I}
{Izidoro} A.,  {Bitsch} B.,  {Raymond} S.~N.,  {Johansen} A.,  {Morbidelli} A.,  {Lambrechts} M.,   {Jacobson} S.~A.,  2021, \mn@doi [\aap] {10.1051/0004-6361/201935336}, 650, A152

\bibitem[\protect\citeauthoryear{{Johansen}, {Davies}, {Church}  \& {Holmelin}}{{Johansen} et~al.}{2012}]{2012ApJ...758...39J}
{Johansen} A.,  {Davies} M.~B.,  {Church} R.~P.,   {Holmelin} V.,  2012, \mn@doi [\apj] {10.1088/0004-637X/758/1/39}, 758, 39

\bibitem[\protect\citeauthoryear{{Koch} et~al.,}{{Koch} et~al.}{2010}]{2010ApJ...713L..79K}
{Koch} D.~G.,  et~al., 2010, \mn@doi [\apjl] {10.1088/2041-8205/713/2/L79}, 713, L79

\bibitem[\protect\citeauthoryear{{Kurosaki} \& {Inutsuka}}{{Kurosaki} \& {Inutsuka}}{2023}]{2023ApJ...954..196K}
{Kurosaki} K.,  {Inutsuka} S.-i.,  2023, \mn@doi [\apj] {10.3847/1538-4357/ace9ba}, 954, 196

\bibitem[\protect\citeauthoryear{{Lada} \& {Lada}}{{Lada} \& {Lada}}{2003}]{RN25}
{Lada} C.~J.,  {Lada} E.~A.,  2003, \mn@doi [\araa] {10.1146/annurev.astro.41.011802.094844}, 41, 57

\bibitem[\protect\citeauthoryear{{Lai} \& {Pu}}{{Lai} \& {Pu}}{2017}]{2017AJ....153...42L}
{Lai} D.,  {Pu} B.,  2017, \mn@doi [\aj] {10.3847/1538-3881/153/1/42}, 153, 42

\bibitem[\protect\citeauthoryear{{Larson}}{{Larson}}{2003}]{2003RPPh...66.1651L}
{Larson} R.~B.,  2003, \mn@doi [Reports on Progress in Physics] {10.1088/0034-4885/66/10/R03}, 66, 1651

\bibitem[\protect\citeauthoryear{{Laughlin} \& {Adams}}{{Laughlin} \& {Adams}}{1998}]{1998ApJ...508L.171L}
{Laughlin} G.,  {Adams} F.~C.,  1998, \mn@doi [\apjl] {10.1086/311736}, 508, L171

\bibitem[\protect\citeauthoryear{{Li}, {Mustill}  \& {Davies}}{{Li} et~al.}{2020}]{2020MNRAS.496.1149L}
{Li} D.,  {Mustill} A.~J.,   {Davies} M.~B.,  2020, \mn@doi [\mnras] {10.1093/mnras/staa1622}, 496, 1149

\bibitem[\protect\citeauthoryear{{Lissauer} et~al.,}{{Lissauer} et~al.}{2011}]{2011ApJS..197....8L}
{Lissauer} J.~J.,  et~al., 2011, \mn@doi [\apjs] {10.1088/0067-0049/197/1/8}, 197, 8

\bibitem[\protect\citeauthoryear{{Liu}, {Hori}, {Lin}  \& {Asphaug}}{{Liu} et~al.}{2015}]{2015ApJ...812..164L}
{Liu} S.-F.,  {Hori} Y.,  {Lin} D.~N.~C.,   {Asphaug} E.,  2015, \mn@doi [\apj] {10.1088/0004-637X/812/2/164}, 812, 164

\bibitem[\protect\citeauthoryear{{Lopez} \& {Fortney}}{{Lopez} \& {Fortney}}{2013}]{2013ApJ...776....2L}
{Lopez} E.~D.,  {Fortney} J.~J.,  2013, \mn@doi [\apj] {10.1088/0004-637X/776/1/2}, 776, 2

\bibitem[\protect\citeauthoryear{{Malmberg}, {Davies}  \& {Heggie}}{{Malmberg} et~al.}{2011}]{2011MNRAS.411..859M}
{Malmberg} D.,  {Davies} M.~B.,   {Heggie} D.~C.,  2011, \mn@doi [\mnras] {10.1111/j.1365-2966.2010.17730.x}, 411, 859

\bibitem[\protect\citeauthoryear{{Marcy}, {Weiss}, {Petigura}, {Isaacson}, {Howard}  \& {Buchhave}}{{Marcy} et~al.}{2014}]{2014PNAS..11112655M}
{Marcy} G.~W.,  {Weiss} L.~M.,  {Petigura} E.~A.,  {Isaacson} H.,  {Howard} A.~W.,   {Buchhave} L.~A.,  2014, \mn@doi [Proceedings of the National Academy of Science] {10.1073/pnas.1304197111}, 111, 12655

\bibitem[\protect\citeauthoryear{{Maschberger}}{{Maschberger}}{2013}]{RN203}
{Maschberger} T.,  2013, \mn@doi [\mnras] {10.1093/mnras/sts479}, 429, 1725

\bibitem[\protect\citeauthoryear{{Millholland}, {Wang}  \& {Laughlin}}{{Millholland} et~al.}{2017}]{2017ApJ...849L..33M}
{Millholland} S.,  {Wang} S.,   {Laughlin} G.,  2017, \mn@doi [\apjl] {10.3847/2041-8213/aa9714}, 849, L33

\bibitem[\protect\citeauthoryear{{Millholland}, {He}, {Ford}, {Ragozzine}, {Fabrycky}  \& {Winn}}{{Millholland} et~al.}{2021}]{2021AJ....162..166M}
{Millholland} S.~C.,  {He} M.~Y.,  {Ford} E.~B.,  {Ragozzine} D.,  {Fabrycky} D.,   {Winn} J.~N.,  2021, \mn@doi [\aj] {10.3847/1538-3881/ac0f7a}, 162, 166

\bibitem[\protect\citeauthoryear{{Moriarty} \& {Ballard}}{{Moriarty} \& {Ballard}}{2016}]{2016ApJ...832...34M}
{Moriarty} J.,  {Ballard} S.,  2016, \mn@doi [\apj] {10.3847/0004-637X/832/1/34}, 832, 34

\bibitem[\protect\citeauthoryear{{Mulders}, {Pascucci}, {Apai}  \& {Ciesla}}{{Mulders} et~al.}{2018}]{2018AJ....156...24M}
{Mulders} G.~D.,  {Pascucci} I.,  {Apai} D.,   {Ciesla} F.~J.,  2018, \mn@doi [\aj] {10.3847/1538-3881/aac5ea}, 156, 24

\bibitem[\protect\citeauthoryear{{Murray} \& {Correia}}{{Murray} \& {Correia}}{2010}]{2010exop.book...15M}
{Murray} C.~D.,  {Correia} A.~C.~M.,  2010, in {Seager} S.,  ed., , Exoplanets.
University of Arizona Press, pp 15--23, \mn@doi{10.48550/arXiv.1009.1738}

\bibitem[\protect\citeauthoryear{{Mustill}, {Davies}  \& {Johansen}}{{Mustill} et~al.}{2017}]{2017MNRAS.468.3000M}
{Mustill} A.~J.,  {Davies} M.~B.,   {Johansen} A.,  2017, \mn@doi [\mnras] {10.1093/mnras/stx693}, 468, 3000

\bibitem[\protect\citeauthoryear{{Paardekooper}, {Dong}, {Duffell}, {Fung}, {Masset}, {Ogilvie}  \& {Tanaka}}{{Paardekooper} et~al.}{2023}]{2023ASPC..534..685P}
{Paardekooper} S.,  {Dong} R.,  {Duffell} P.,  {Fung} J.,  {Masset} F.~S.,  {Ogilvie} G.,   {Tanaka} H.,  2023, in {Inutsuka} S.,  {Aikawa} Y.,  {Muto} T.,  {Tomida} K.,   {Tamura} M.,  eds,  Astronomical Society of the Pacific Conference Series Vol. 534, Astronomical Society of the Pacific Conference Series. p.~685

\bibitem[\protect\citeauthoryear{{Parker}}{{Parker}}{2020}]{2020RSOS....701271P}
{Parker} R.~J.,  2020, \mn@doi [Royal Society Open Science] {10.1098/rsos.201271}, 7, 201271

\bibitem[\protect\citeauthoryear{{Parker} \& {Quanz}}{{Parker} \& {Quanz}}{2012}]{2012MNRAS.419.2448P}
{Parker} R.~J.,  {Quanz} S.~P.,  2012, \mn@doi [\mnras] {10.1111/j.1365-2966.2011.19911.x}, 419, 2448

\bibitem[\protect\citeauthoryear{{Parker} \& {Wright}}{{Parker} \& {Wright}}{2016}]{RN1}
{Parker} R.~J.,  {Wright} N.~J.,  2016, \mn@doi [\mnras] {10.1093/mnras/stw087}, 457, 3430

\bibitem[\protect\citeauthoryear{{Parker}, {Wright}, {Goodwin}  \& {Meyer}}{{Parker} et~al.}{2014}]{RN5}
{Parker} R.~J.,  {Wright} N.~J.,  {Goodwin} S.~P.,   {Meyer} M.~R.,  2014, \mn@doi [\mnras] {10.1093/mnras/stt2231}, 438, 620

\bibitem[\protect\citeauthoryear{{Petigura}, {Howard}  \& {Marcy}}{{Petigura} et~al.}{2013}]{2013PNAS..11019273P}
{Petigura} E.~A.,  {Howard} A.~W.,   {Marcy} G.~W.,  2013, \mn@doi [Proceedings of the National Academy of Science] {10.1073/pnas.1319909110}, 110, 19273

\bibitem[\protect\citeauthoryear{{Petigura} et~al.,}{{Petigura} et~al.}{2022}]{Petigura2022}
{Petigura} E.~A.,  et~al., 2022, \mn@doi [\aj] {10.3847/1538-3881/ac51e3}, 163, 179

\bibitem[\protect\citeauthoryear{{Pfalzner}, {Bhandare}, {Vincke}  \& {Lacerda}}{{Pfalzner} et~al.}{2018}]{2018ApJ...863...45P}
{Pfalzner} S.,  {Bhandare} A.,  {Vincke} K.,   {Lacerda} P.,  2018, \mn@doi [\apj] {10.3847/1538-4357/aad23c}, 863, 45

\bibitem[\protect\citeauthoryear{{Pfalzner}, {Aizpuru Vargas}, {Bhandare}  \& {Veras}}{{Pfalzner} et~al.}{2021}]{2021A&A...651A..38P}
{Pfalzner} S.,  {Aizpuru Vargas} L.~L.,  {Bhandare} A.,   {Veras} D.,  2021, \mn@doi [\aap] {10.1051/0004-6361/202140587}, 651, A38

\bibitem[\protect\citeauthoryear{{Picogna} \& {Marzari}}{{Picogna} \& {Marzari}}{2014}]{2014A&A...564A..28P}
{Picogna} G.,  {Marzari} F.,  2014, \mn@doi [\aap] {10.1051/0004-6361/201322816}, 564, A28

\bibitem[\protect\citeauthoryear{{Plummer}}{{Plummer}}{1911}]{RN198}
{Plummer} H.~C.,  1911, \mn@doi [\mnras] {10.1093/mnras/71.5.460}, 71, 460

\bibitem[\protect\citeauthoryear{{Poon} \& {Nelson}}{{Poon} \& {Nelson}}{2020}]{2020MNRAS.498.5166P}
{Poon} S. T.~S.,  {Nelson} R.~P.,  2020, \mn@doi [\mnras] {10.1093/mnras/staa2755}, 498, 5166

\bibitem[\protect\citeauthoryear{{Pu} \& {Lai}}{{Pu} \& {Lai}}{2018}]{2018MNRAS.478..197P}
{Pu} B.,  {Lai} D.,  2018, \mn@doi [\mnras] {10.1093/mnras/sty1098}, 478, 197

\bibitem[\protect\citeauthoryear{{Pu} \& {Wu}}{{Pu} \& {Wu}}{2015}]{2015ApJ...807...44P}
{Pu} B.,  {Wu} Y.,  2015, \mn@doi [\apj] {10.1088/0004-637X/807/1/44}, 807, 44

\bibitem[\protect\citeauthoryear{{Ragozzine} \& {Holman}}{{Ragozzine} \& {Holman}}{2010}]{2010arXiv1006.3727R}
{Ragozzine} D.,  {Holman} M.~J.,  2010, \mn@doi [arXiv e-prints] {10.48550/arXiv.1006.3727}, p. arXiv:1006.3727

\bibitem[\protect\citeauthoryear{{Read}, {Wyatt}  \& {Triaud}}{{Read} et~al.}{2017}]{2017MNRAS.469..171R}
{Read} M.~J.,  {Wyatt} M.~C.,   {Triaud} A. H.~M.~J.,  2017, \mn@doi [\mnras] {10.1093/mnras/stx798}, 469, 171

\bibitem[\protect\citeauthoryear{{Rein} \& {Liu}}{{Rein} \& {Liu}}{2012}]{rebound}
{Rein} H.,  {Liu} S.~F.,  2012, \mn@doi [\aap] {10.1051/0004-6361/201118085}, 537, A128

\bibitem[\protect\citeauthoryear{{Rein} \& {Spiegel}}{{Rein} \& {Spiegel}}{2015}]{reboundias15}
{Rein} H.,  {Spiegel} D.~S.,  2015, \mn@doi [\mnras] {10.1093/mnras/stu2164}, 446, 1424

\bibitem[\protect\citeauthoryear{{Rein} \& {Tamayo}}{{Rein} \& {Tamayo}}{2015}]{2015MNRAS.452..376R}
{Rein} H.,  {Tamayo} D.,  2015, \mn@doi [\mnras] {10.1093/mnras/stv1257}, 452, 376

\bibitem[\protect\citeauthoryear{{Rein} \& {Tamayo}}{{Rein} \& {Tamayo}}{2017}]{reboundsa}
{Rein} H.,  {Tamayo} D.,  2017, \mn@doi [\mnras] {10.1093/mnras/stx232}, 467, 2377

\bibitem[\protect\citeauthoryear{{Rein} et~al.,}{{Rein} et~al.}{2019}]{reboundmercurius}
{Rein} H.,  et~al., 2019, \mn@doi [\mnras] {10.1093/mnras/stz769}, 485, 5490

\bibitem[\protect\citeauthoryear{{Rickman}, {Wajer}, {Przy{\l}uski}, {Wi{\'s}niowski}, {Nesvorn{\'y}}  \& {Morbidelli}}{{Rickman} et~al.}{2023}]{2023MNRAS.520..637R}
{Rickman} H.,  {Wajer} P.,  {Przy{\l}uski} R.,  {Wi{\'s}niowski} T.,  {Nesvorn{\'y}} D.,   {Morbidelli} A.,  2023, \mn@doi [\mnras] {10.1093/mnras/stac3705}, 520, 637

\bibitem[\protect\citeauthoryear{{Rodet} \& {Lai}}{{Rodet} \& {Lai}}{2022}]{2022MNRAS.509.1010R}
{Rodet} L.,  {Lai} D.,  2022, \mn@doi [\mnras] {10.1093/mnras/stab3046}, 509, 1010

\bibitem[\protect\citeauthoryear{{Rogers} \& {Owen}}{{Rogers} \& {Owen}}{2021}]{2021MNRAS.503.1526R}
{Rogers} J.~G.,  {Owen} J.~E.,  2021, \mn@doi [\mnras] {10.1093/mnras/stab529}, 503, 1526

\bibitem[\protect\citeauthoryear{{Rosenthal} et~al.,}{{Rosenthal} et~al.}{2022}]{2022ApJS..262....1R}
{Rosenthal} L.~J.,  et~al., 2022, \mn@doi [\apjs] {10.3847/1538-4365/ac7230}, 262, 1

\bibitem[\protect\citeauthoryear{{Safronov}}{{Safronov}}{1972}]{1972epcf.book.....S}
{Safronov} V.~S.,  1972, {Evolution of the protoplanetary cloud and formation of the earth and planets.}.
Keter Publishing House

\bibitem[\protect\citeauthoryear{{Salpeter}}{{Salpeter}}{1955}]{RN204}
{Salpeter} E.~E.,  1955, \mn@doi [\apj] {10.1086/145971}, 121, 161

\bibitem[\protect\citeauthoryear{{Schlecker}, {Mordasini}, {Emsenhuber}, {Klahr}, {Henning}, {Burn}, {Alibert}  \& {Benz}}{{Schlecker} et~al.}{2021}]{2021A&A...656A..71S}
{Schlecker} M.,  {Mordasini} C.,  {Emsenhuber} A.,  {Klahr} H.,  {Henning} T.,  {Burn} R.,  {Alibert} Y.,   {Benz} W.,  2021, \mn@doi [\aap] {10.1051/0004-6361/202038554}, 656, A71

\bibitem[\protect\citeauthoryear{{Schoettler}, {de Bruijne}, {Vaher}  \& {Parker}}{{Schoettler} et~al.}{2020}]{2020MNRAS.495.3104S}
{Schoettler} C.,  {de Bruijne} J.,  {Vaher} E.,   {Parker} R.~J.,  2020, \mn@doi [\mnras] {10.1093/mnras/staa1228}, 495, 3104

\bibitem[\protect\citeauthoryear{{Schoettler}, {Parker}  \& {de Bruijne}}{{Schoettler} et~al.}{2022}]{2022MNRAS.510.3178S}
{Schoettler} C.,  {Parker} R.~J.,   {de Bruijne} J.,  2022, \mn@doi [\mnras] {10.1093/mnras/stab3529}, 510, 3178

\bibitem[\protect\citeauthoryear{{Segura-Cox} et~al.,}{{Segura-Cox} et~al.}{2020}]{2020Natur.586..228S}
{Segura-Cox} D.~M.,  et~al., 2020, \mn@doi [\nat] {10.1038/s41586-020-2779-6}, 586, 228

\bibitem[\protect\citeauthoryear{{Shara}, {Hurley}  \& {Mardling}}{{Shara} et~al.}{2016}]{2016ApJ...816...59S}
{Shara} M.~M.,  {Hurley} J.~R.,   {Mardling} R.~A.,  2016, \mn@doi [\apj] {10.3847/0004-637X/816/2/59}, 816, 59

\bibitem[\protect\citeauthoryear{{Spalding} \& {Batygin}}{{Spalding} \& {Batygin}}{2016}]{2016ApJ...830....5S}
{Spalding} C.,  {Batygin} K.,  2016, \mn@doi [\apj] {10.3847/0004-637X/830/1/5}, 830, 5

\bibitem[\protect\citeauthoryear{{Spurzem}, {Giersz}, {Heggie}  \& {Lin}}{{Spurzem} et~al.}{2009}]{2009ApJ...697..458S}
{Spurzem} R.,  {Giersz} M.,  {Heggie} D.~C.,   {Lin} D.~N.~C.,  2009, \mn@doi [\apj] {10.1088/0004-637X/697/1/458}, 697, 458

\bibitem[\protect\citeauthoryear{{Stock}, {Cai}, {Spurzem}, {Kouwenhoven}  \& {Portegies Zwart}}{{Stock} et~al.}{2020}]{2020MNRAS.497.1807S}
{Stock} K.,  {Cai} M.~X.,  {Spurzem} R.,  {Kouwenhoven} M.~B.~N.,   {Portegies Zwart} S.,  2020, \mn@doi [\mnras] {10.1093/mnras/staa2047}, 497, 1807

\bibitem[\protect\citeauthoryear{{Stock}, {Veras}, {Cai}, {Spurzem}  \& {Portegies Zwart}}{{Stock} et~al.}{2022}]{2022MNRAS.512.2460S}
{Stock} K.,  {Veras} D.,  {Cai} M.~X.,  {Spurzem} R.,   {Portegies Zwart} S.,  2022, \mn@doi [\mnras] {10.1093/mnras/stac602}, 512, 2460

\bibitem[\protect\citeauthoryear{{Thorngren}, {Marley}  \& {Fortney}}{{Thorngren} et~al.}{2019}]{2019RNAAS...3..128T}
{Thorngren} D.~P.,  {Marley} M.~S.,   {Fortney} J.~J.,  2019, \mn@doi [Research Notes of the American Astronomical Society] {10.3847/2515-5172/ab4353}, 3, 128

\bibitem[\protect\citeauthoryear{{Van Eylen} \& {Albrecht}}{{Van Eylen} \& {Albrecht}}{2015}]{2015ApJ...808..126V}
{Van Eylen} V.,  {Albrecht} S.,  2015, \mn@doi [\apj] {10.1088/0004-637X/808/2/126}, 808, 126

\bibitem[\protect\citeauthoryear{{Wang}, {Kanagawa}, {Hayashi}  \& {Suto}}{{Wang} et~al.}{2020a}]{2020ApJ...891..166W}
{Wang} S.,  {Kanagawa} K.~D.,  {Hayashi} T.,   {Suto} Y.,  2020a, \mn@doi [\apj] {10.3847/1538-4357/ab781b}, 891, 166

\bibitem[\protect\citeauthoryear{{Wang}, {Leigh}, {Perna}  \& {Shara}}{{Wang} et~al.}{2020b}]{2020ApJ...905..136W}
{Wang} Y.-H.,  {Leigh} N. W.~C.,  {Perna} R.,   {Shara} M.~M.,  2020b, \mn@doi [\apj] {10.3847/1538-4357/abc619}, 905, 136

\bibitem[\protect\citeauthoryear{{Wang}, {Perna}, {Leigh}  \& {Shara}}{{Wang} et~al.}{2022}]{2022MNRAS.509.5253W}
{Wang} Y.-H.,  {Perna} R.,  {Leigh} N. W.~C.,   {Shara} M.~M.,  2022, \mn@doi [\mnras] {10.1093/mnras/stab3321}, 509, 5253

\bibitem[\protect\citeauthoryear{{Weiss} et~al.,}{{Weiss} et~al.}{2018}]{2018AJ....155...48W}
{Weiss} L.~M.,  et~al., 2018, \mn@doi [\aj] {10.3847/1538-3881/aa9ff6}, 155, 48

\bibitem[\protect\citeauthoryear{{Weiss}, {Millholland}, {Petigura}, {Adams}, {Batygin}, {Bloch}  \& {Mordasini}}{{Weiss} et~al.}{2022}]{2022arXiv220310076W}
{Weiss} L.~M.,  {Millholland} S.~C.,  {Petigura} E.~A.,  {Adams} F.~C.,  {Batygin} K.,  {Bloch} A.~M.,   {Mordasini} C.,  2022, \mn@doi [arXiv e-prints] {10.48550/arXiv.2203.10076}, p. arXiv:2203.10076

\bibitem[\protect\citeauthoryear{{Winn}}{{Winn}}{2010}]{2010arXiv1001.2010W}
{Winn} J.~N.,  2010, \mn@doi [arXiv e-prints] {10.48550/arXiv.1001.2010}, p. arXiv:1001.2010

\bibitem[\protect\citeauthoryear{{Xie} et~al.,}{{Xie} et~al.}{2016}]{2016PNAS..11311431X}
{Xie} J.-W.,  et~al., 2016, \mn@doi [Proceedings of the National Academy of Science] {10.1073/pnas.1604692113}, 113, 11431

\bibitem[\protect\citeauthoryear{{Zawadzki}, {Carrera}  \& {Ford}}{{Zawadzki} et~al.}{2022}]{2022ApJ...937...53Z}
{Zawadzki} B.,  {Carrera} D.,   {Ford} E.~B.,  2022, \mn@doi [\apj] {10.3847/1538-4357/ac8b04}, 937, 53

\bibitem[\protect\citeauthoryear{{Zhu} \& {Wu}}{{Zhu} \& {Wu}}{2018}]{2018AJ....156...92Z}
{Zhu} W.,  {Wu} Y.,  2018, \mn@doi [\aj] {10.3847/1538-3881/aad22a}, 156, 92

\bibitem[\protect\citeauthoryear{{Zhu}, {Petrovich}, {Wu}, {Dong}  \& {Xie}}{{Zhu} et~al.}{2018}]{2018ApJ...860..101Z}
{Zhu} W.,  {Petrovich} C.,  {Wu} Y.,  {Dong} S.,   {Xie} J.,  2018, \mn@doi [\apj] {10.3847/1538-4357/aac6d5}, 860, 101

\bibitem[\protect\citeauthoryear{{Zink}, {Christiansen}  \& {Hansen}}{{Zink} et~al.}{2019}]{2019MNRAS.483.4479Z}
{Zink} J.~K.,  {Christiansen} J.~L.,   {Hansen} B. M.~S.,  2019, \mn@doi [\mnras] {10.1093/mnras/sty3463}, 483, 4479

\bibitem[\protect\citeauthoryear{{van Elteren}, {Portegies Zwart}, {Pelupessy}, {Cai}  \& {McMillan}}{{van Elteren} et~al.}{2019}]{2019A&A...624A.120V}
{van Elteren} A.,  {Portegies Zwart} S.,  {Pelupessy} I.,  {Cai} M.~X.,   {McMillan} S.~L.~W.,  2019, \mn@doi [\aap] {10.1051/0004-6361/201834641}, 624, A120

\makeatother
\end{thebibliography}



\appendix

\textcolor{changes}{\section{Effect of fly-by velocity on mutual inclination evolution of inner planets}}
\begin{figure*}
    \centering
    \begin{minipage}[t]{1.0\columnwidth}
         \centering
    	\includegraphics[width=1.0\linewidth]{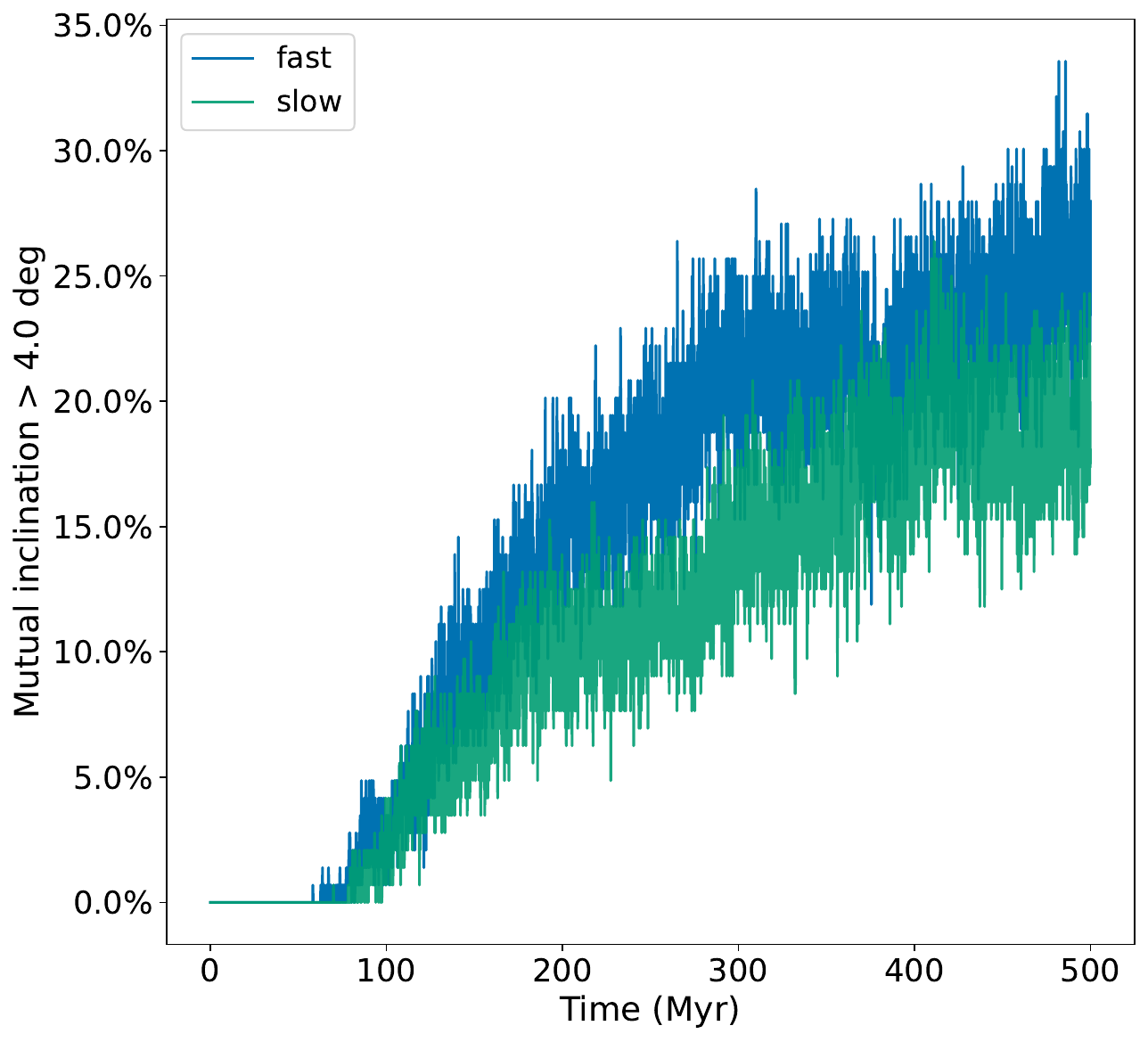}
     \end{minipage}
        \centering
        \vspace{0pt}
    \begin{minipage}[t]{1.0\columnwidth}
    	\includegraphics[width=1\linewidth]{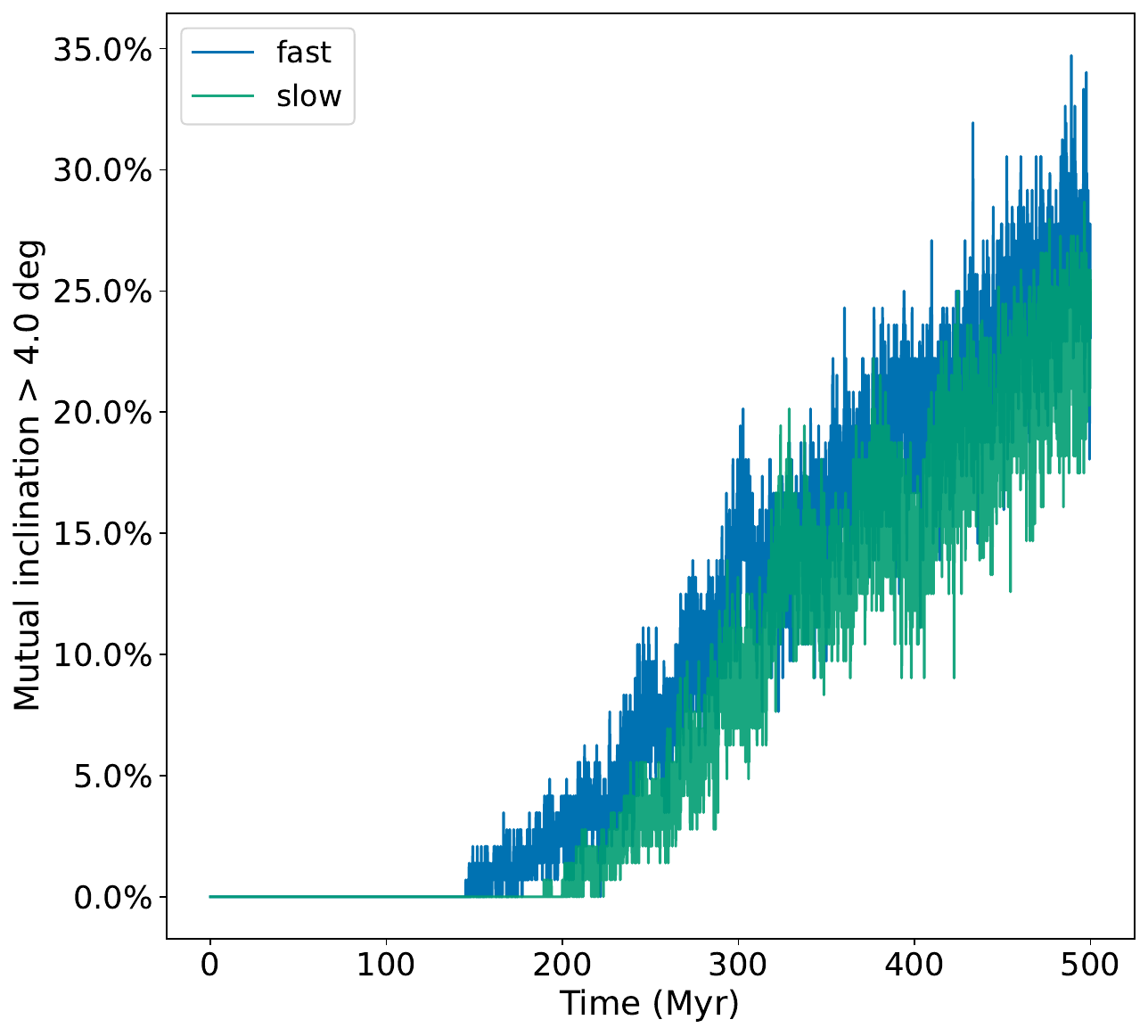}
      \end{minipage}
     \caption{\textcolor{changes}{Mutual inclination evolution between the inner planets for 10 $R_{\rm{MH}}$ (left) and 14 $R_{\rm{MH}}$ (right) initial separation. Evolution of the fraction of simulations that show a maximum mutual inclination above 4.0$\degree$ for the two different fly-by velocities used in the simulations. The 'fast' ('blue') evolution is for those with 4 km\,s$^{-1}$ velocity, compared to the 'slow' ('green') fly-bys with 2 km\,s$^{-1}$ velocity. The simulations with the slower velocities will reach a similar level of perturbation as evidenced by the maximum mutual inclinations than the faster velocity simulations but slightly offset in time.}}
     \label{fig:Mut_inc_vel_evolv}
\end{figure*}


\bsp	
\label{lastpage}
\end{document}